\documentclass[manuscript=article]{achemso}
\setkeys{acs}{articletitle = true}
\usepackage[T1]{fontenc}
\usepackage[utf8]{inputenc}
\usepackage{graphicx}
\usepackage{amsmath}
\usepackage{color}
\usepackage{comment}
\usepackage{braket}
\usepackage{hyperref}
\usepackage{multirow}
\usepackage{subfig}
\usepackage{amsmath,mathtools}
\usepackage[version=3]{mhchem}
\usepackage{longtable}
\usepackage{siunitx}
\usepackage{tablefootnote}
\usepackage[normalem]{ulem}
\usepackage[para,online]{threeparttable}

\definecolor{ddc}{rgb}{0.5, 0.0, 0.5}

\definecolor{ao}{rgb}{0.0, 0.5, 0.0}
\definecolor{bs}{rgb}{1.0, 0.44, 0.37}
\definecolor{magenta}{rgb}{1.0, 0.0, 1.0}

\title{
The role of spin polarization and dynamic correlation in singlet-triplet gap inversion of heptazine derivatives}
\author{Daria Drwal}
\affiliation{Institute of Physics, Lodz University of Technology, \mbox{ul.\ Wolczanska 219, 90-924 Lodz, Poland}}
\altaffiliation{Contributed equally.}

\author{Mikulas Matousek}
\affiliation{J. Heyrovsk\'{y} Institute of Physical Chemistry, Academy of Sciences of the Czech \mbox{Republic, v.v.i.}, Dolej\v{s}kova 3, 18223 Prague 8, Czech Republic}
\alsoaffiliation{Faculty of Mathematics and Physics, Charles University, Prague, Czech Republic}
\altaffiliation{Contributed equally.}

\author{Pavlo Golub}
\affiliation{J. Heyrovsk\'{y} Institute of Physical Chemistry, Academy of Sciences of the Czech \mbox{Republic, v.v.i.}, Dolej\v{s}kova 3, 18223 Prague 8, Czech Republic}

\author{Aleksandra Tucholska}
\affiliation{Institute of Physics, Lodz University of Technology, \mbox{ul.\ Wolczanska 219, 90-924 Lodz, Poland}}

\author{Michał Hapka}
\affiliation{Faculty of Chemistry, University of Warsaw, ul.\ L.\ Pasteura 1, 02-093 Warsaw, Poland}

\author{Jiri Brabec}
\affiliation{J. Heyrovsk\'{y} Institute of Physical Chemistry, Academy of Sciences of the Czech \mbox{Republic, v.v.i.}, Dolej\v{s}kova 3, 18223 Prague 8, Czech Republic}

\author{Libor Veis}
\email{libor.veis@jh-inst.cas.cz}
\affiliation{J. Heyrovsk\'{y} Institute of Physical Chemistry, Academy of Sciences of the Czech \mbox{Republic, v.v.i.}, Dolej\v{s}kova 3, 18223 Prague 8, Czech Republic}

\author{Katarzyna Pernal}
\email{pernalk@gmail.com}
\affiliation{Institute of Physics, Lodz University of Technology, \mbox{ul.\ Wolczanska 219, 90-924 Lodz, Poland}}


\begin{document}

\begin{abstract}
    The new generation of proposed light-emitting molecules for OLEDs has raised a considerable research interest due to its exceptional feature---a negative singlet-triplet (ST) gap  violating the Hund’s multiplicity rule in the excited S$_1$ and T$_1$ states. We investigate the role of spin polarization in the mechanism of ST gap inversion. Spin polarization is associated with doubly excited determinants of certain types, whose presence in the wavefunction expansion favors the energy of the singlet state more than that of the triplet. 
    Using a perturbation theory-based model for spin polarization, we propose a simple descriptor for prescreening of candidate molecules with negative ST gaps and prove its usefulness for heptazine-type molecules. Numerical results show that the quantitative effect of spin polarization is approximately inverse-proportional to the HOMO-LUMO exchange integral. Comparison of single- and multireference coupled-cluster predictions of ST gaps shows that the former methods provide good accuracy by correctly balancing the effects of doubly excited determinants and dynamic correlation. We also show that accurate ST gaps may be obtained using a complete active space model supplemented with dynamic correlation from multireference adiabatic connection theory.
\end{abstract}  

\section{Introduction}

 Detecting candidates for organic light-emitting diodes (OLEDs) through screening of  potential chromophores remains a challenge for quantum chemistry methods. OLED molecules may rely on different light-emitting mechanisms, ranging from pure fluorescence or phosphorescence (first and second generation emitters) to more complex processes aimed at harvesting nonfluorescent triplet excitons (third and fourth generation).~\cite{tyan2011organic,Hong:21} 
Typically,  organic molecules obey Hund’s multiplicity rule\cite{hund1925deutung} and the  triplet state T$_1$ is lower than the first excited singlet state S$_1$. Hence, only a minor part of the excitons is available for photon emission through radiative recombination of singlets. This restricts internal quantum efficiency (IQE) to 25\% when the distribution of excitons between S$_1$ and T$_1$ states is most favorable \cite{4_li2006organic}.
For molecules characterized by small singlet-triplet gaps it is possible to thermally induce reverse intersystem crossing (RISC), i.e.\ transition from lower-lying T$_1$ (dark state) to higher S$_1$ (bright state). This process,  known as thermally activated delayed fluorescence (TADF), results in higher-rate fluorescent emission from the singlet state.~\cite{1SD2_endo2011efficient,2SD2_uoyama2012highly,3SD2_goushi2012organic,4SD2_nakanotani2014high,5SD2_hosokai2017evidence}

Recent attention has focused on systems with inverted singlet-triplet gap (INVEST) molecules. In INVEST emitters the S$_1$ state lies below T$_1$, so that the relaxation from T$_1$ takes place with no need for thermal induction and 100\% IQE of luminescence could, in principle, be achieved.~\cite{desilva2019inverted,sobolewskidomcke2021exp,sobolewskidomcke2021,ricci2022establishing}
It is worth mentioning that aromatic chromophores  with negative S$_1-$T$_1$ energy difference (ST gap) are of interest in another area of active research, namely in water-splitting photocatalysis. It has been shown that derivatives of heptazine may act as efficient photocatalysts. Thanks to the negative ST energy gap, the S$_1$ state is exceptionally long-lived, as it does not suffer from quenching through intersystem crossing (ISC) to T$_1$.\cite{kataliza_ehrmaier2020molecular} 

First calculations proving the possibility of ST gap inversion were carried out independently in 2019 by de Silva for cycl[3.3.3]azine\cite{desilva2019inverted} and Sobolewski, Domcke and co-workers for heptazine.~\cite{6fun_ehrmaier2019singlet} Ref.~\citenum{6fun_ehrmaier2019singlet} presented also the first indirect experimental proof for the gap inversion and indicated that this phenomenon could explain the high efficiency of OLEDs observed already in 2013\cite{Adachi:13} and 2014\cite{33SD2_li2014thermally,35SD2_li2014highly} by Adachi and co-workers. A subsequent theoretical study by Sobolewski and Domcke\cite{sobolewskidomcke2021exp} confirmed this hypothesis. Works of Sancho-Garcia and co-workers\cite{27SD2_ricci2021singlet,28SD2_sanz2021negative} demonstrated ST gap inversion across N- and B-doped triangulenes of different sizes. In 2022, Aizawa and co-workers\cite{nature_aizawa2022delayed} proposed a prescreening approach for INVEST candidates. Based on the elimination process, two heptazine analogues were chosen for further, experimental evaluation, which confirmed gap inversion.

The magnitude of the S$_1$-T$_1$ energy gap is directly related to the exchange integral involving the highest occupied molecular orbital (HOMO) and lowest unoccupied molecular orbital (LUMO).~\cite{staemmler1978violation,guzik2021organic,sundholm}  
A vanishing overlap between frontier orbitals is a prerequisite for ST inversion, but it is not sufficient.
Several studies have demonstrated that accounting for electron correlation via inclusion of double excitations is essential to stabilize the singlet state with respect to triplet and obtain  negative ST gaps of chromophore molecules.~\cite{desilva2019inverted,sundholm,sobolewskidomcke2021} Results from single-reference response methods suggest that double excitations contribute at a relatively low level to the S$_1$ state, estimated at ca.\ 10\% by de Silva.~\cite{desilva2019inverted} Time-dependent density functional theory (TD-DFT) approaches are incapable of capturing   double excitations, and do not predict negative gaps, as corroborated by a range of investigations.~\cite{desilva2019inverted,6fun_ehrmaier2019singlet,7fun_ricci2021singlet,8fun_sanz2021negative,9fun_bhattacharyya2021can,10fun_ghosh2022origin}

In pursuit of efficient strategies in designing novel INVEST materials, Pollice et al.\cite{guzik2021organic}  investigated all possible permutations of cyclazine and heptazine derivatives with C--H replaced by a set of electron-donating and electron-withdrawing substituents, and described them at the EOM-CCSD and TD-DFT levels of theory. 
Aizawa and coworkers\cite{nature_aizawa2022delayed} expanded on this work by introducing 186 new substituents and finding almost 35 000 potential candidates for INVEST molecules. Their further TD-DFT screening resulted in a significant limitation of this set.  Still, TD-DFT-based screening is inherently limited, as the method  cannot distinguish INVEST molecules from cases in which the ST gap is small, but not negative.
Having studied the relationship between the structure and properties of selected INVEST molecules, Olivier and co-workers\cite{ricci2022establishing} formulated a set of design rules in which the $C_{\rm 2v}$ point group of the triangulene core was identified as a prerequisite for the gap inversion.

Advances in designing novel INVEST systems are hindered by two factors: limited understanding of the mechanism behind gap reversal at the electronic structure level and the lack of efficient descriptors  that could account for the main effects lowering S$_1$ below T$_1$. 
In one of the first theoretical works devoted to the violation of Hund's rule in closed-shell molecules, Kollmar and Staemmler\cite{staemmler1978violation} introduced a concept of dynamic spin polarization (sp), which associates the ST gap inversion to the energetic effect exerted by a small subset of doubly-excited configurations involving frontier orbitals. Numerical investigations on conjugated hydrocarbons carried out by Koseki et al.\cite{koseki1985violation} showed that spin polarization may indeed lead to gap inversion in some molecules. Spin polarization has been also indirectly identified as a key effect in ST gap inversion of the heptazine molecule, without, however, providing its quantitative measure.\cite{6fun_ehrmaier2019singlet}

The main objective of our study is to elucidate the role of both spin polarization and dynamic correlation energy in the mechanism of ST gap inversion in heptazine-based molecules. 
Based on the spin polarization model, we propose a simple descriptor for prescreening of molecules with a negative ST gap. We also demonstrate the efficacy of selected multireference methods in accurately predicting the magnitude of ST gaps in organic INVEST emitters.  We compare multireference adiabatic connection methods\cite{harris,perdew,gunnarsson,helgaker_ac,ac_prl,Pernal:18b,Drwal:22} and the second-order $n$-electron valence state perturbation theory (NEVPT2)\cite{nevpt2_from_biradicals} against Mukherjee{\textquotesingle}s multireference coupled-cluster with non-iterative triple excitations [Mk-MRCCSD(T)].~\cite{mrcc}

\section{Computational details}

\begin{figure}[ht]
\centering
\begin{minipage}[b]{0.32\textwidth}
\includegraphics[scale=0.1]{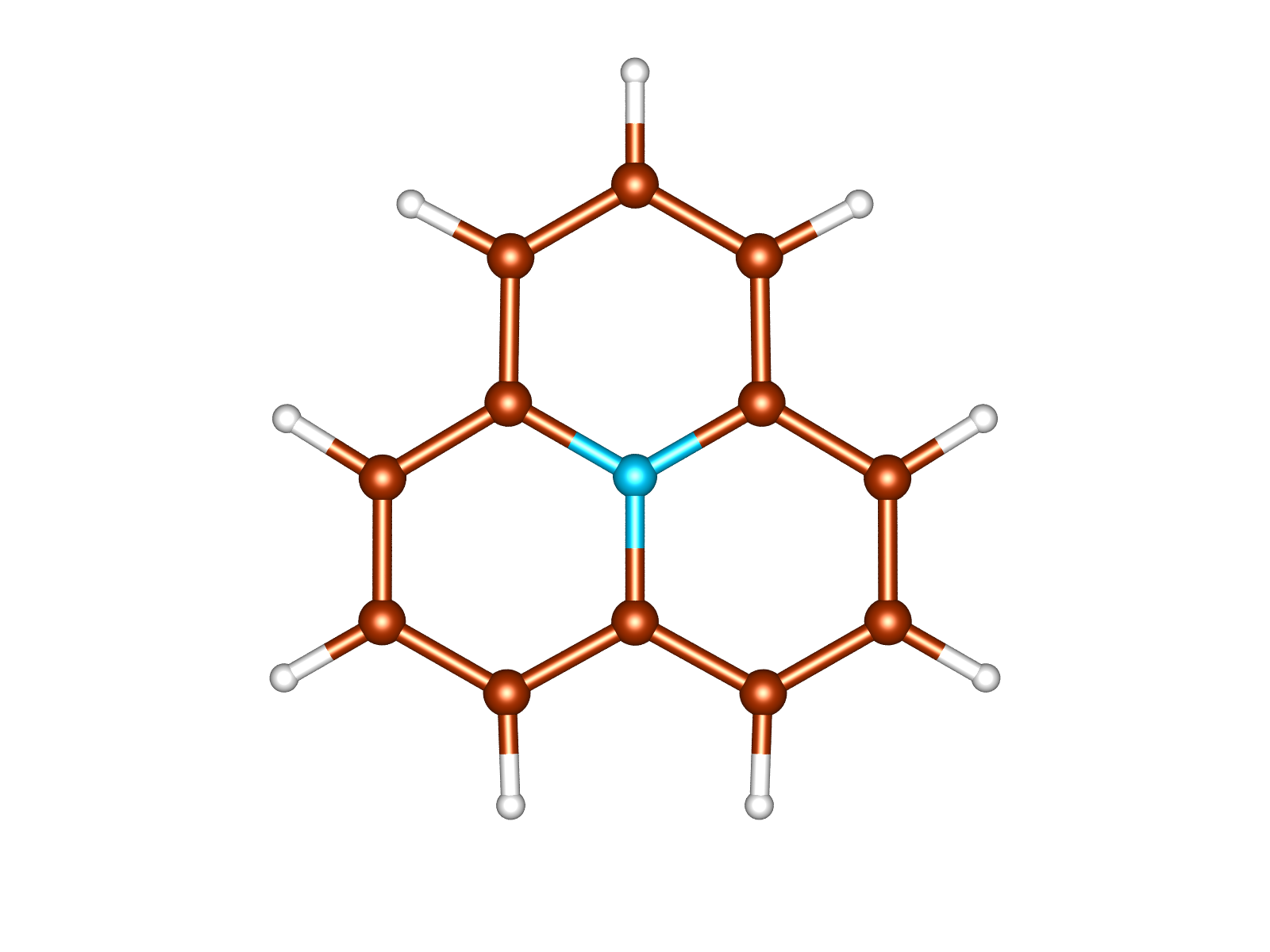}
\centering
\textbf{1}
\end{minipage}
\begin{minipage}[b]{0.32\textwidth}
\includegraphics[scale=0.1]{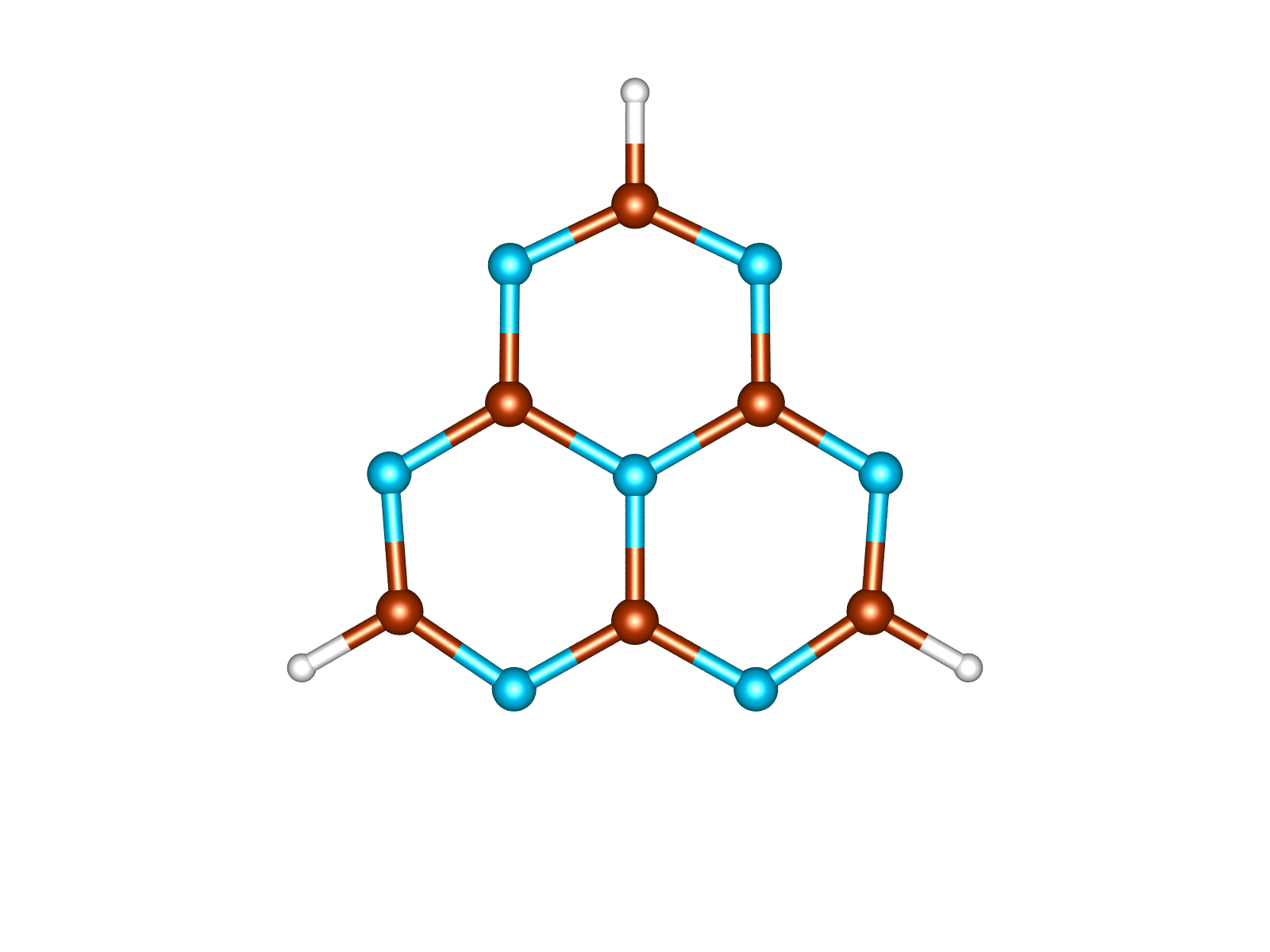}
\centering
\textbf{2}
\end{minipage}
\begin{minipage}[b]{0.32\textwidth}
\includegraphics[scale=0.1]{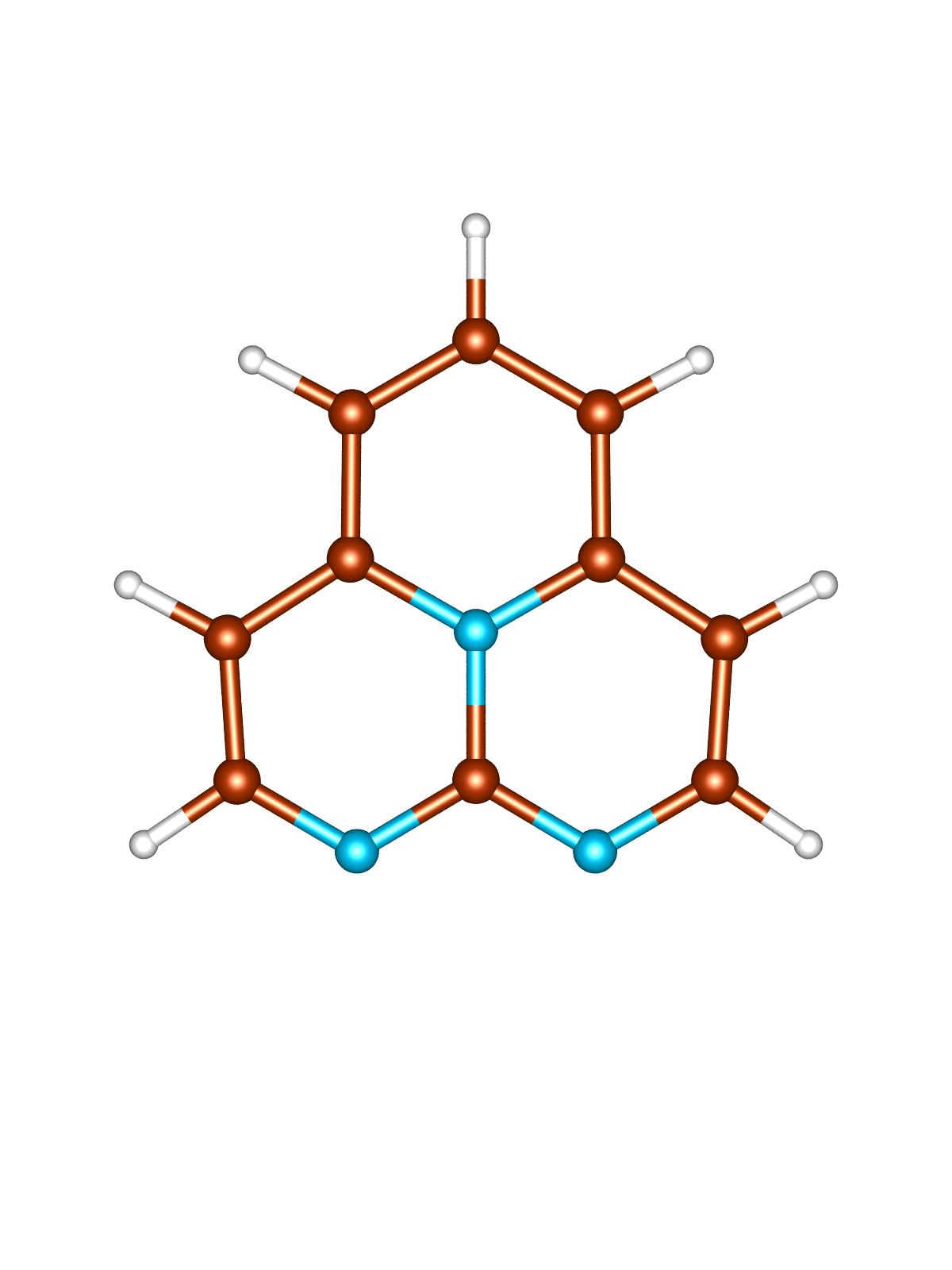}
\centering
\textbf{3}
\end{minipage}
\begin{minipage}[b]{0.32\textwidth}
\includegraphics[scale=0.1]{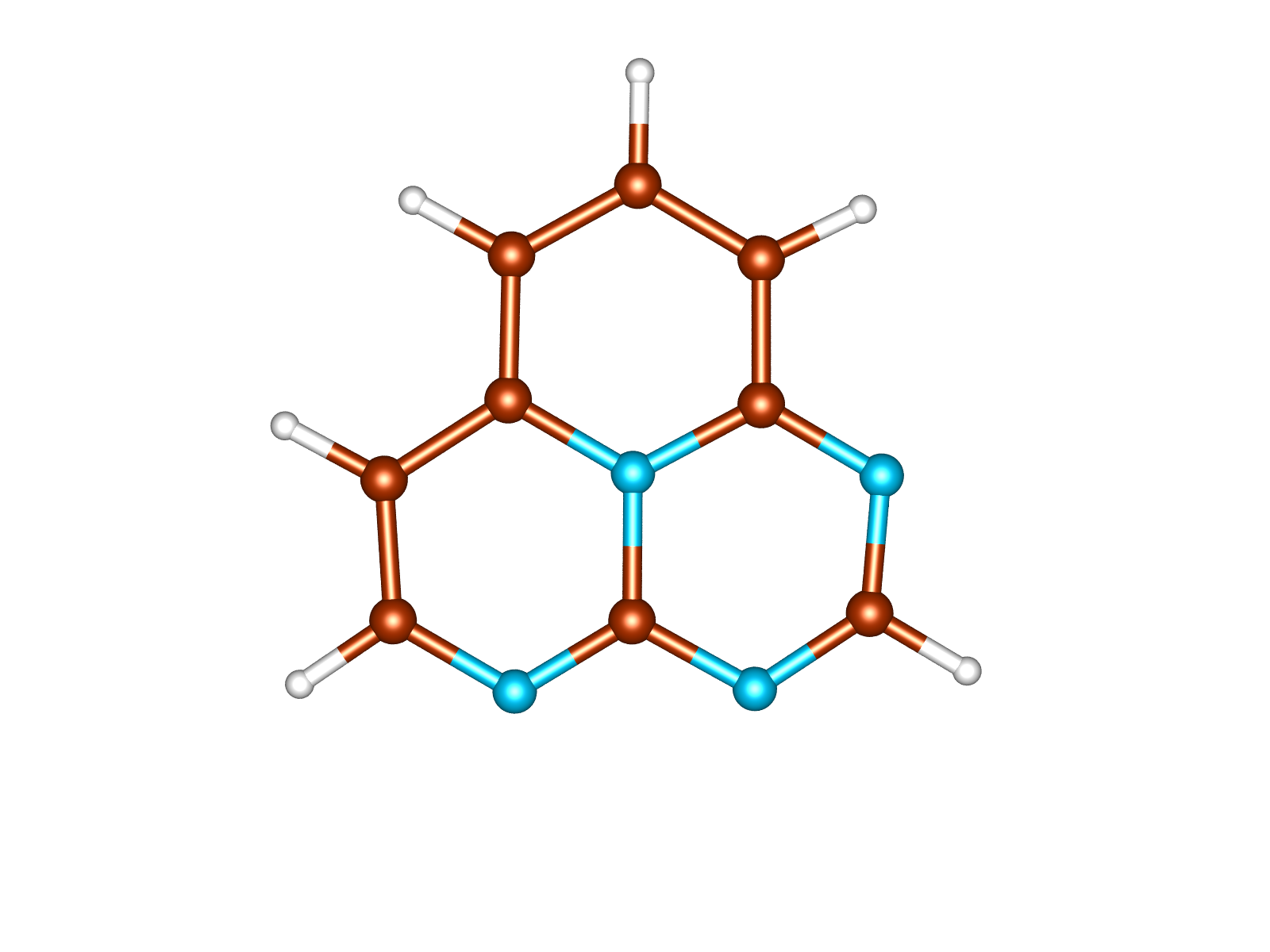}
\centering
\textbf{4}
\end{minipage}
\begin{minipage}[b]{0.32\textwidth}
\includegraphics[scale=0.1]{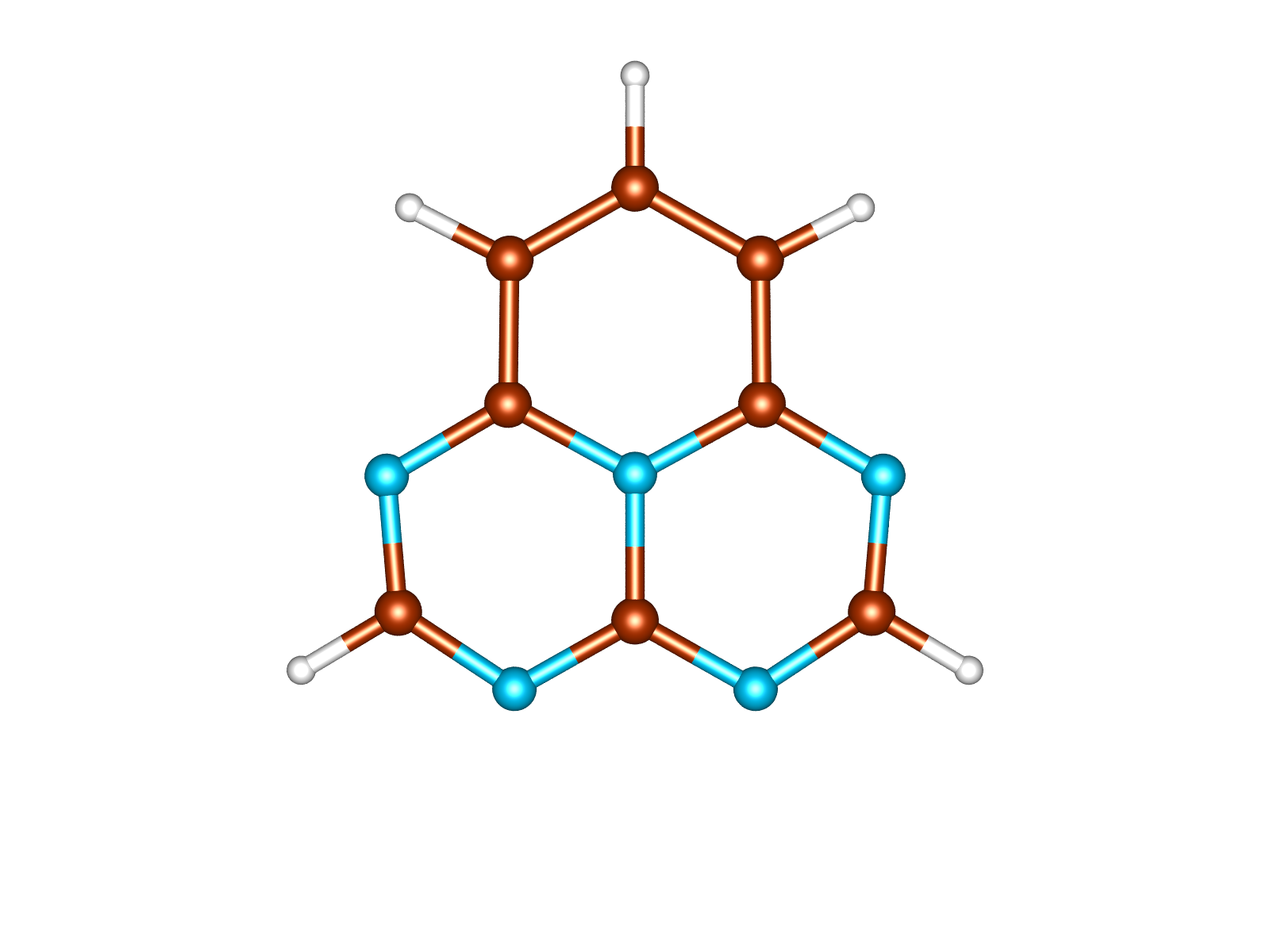}
\centering
\textbf{5}
\end{minipage}
\begin{minipage}[b]{0.32\textwidth}
\includegraphics[scale=0.1]{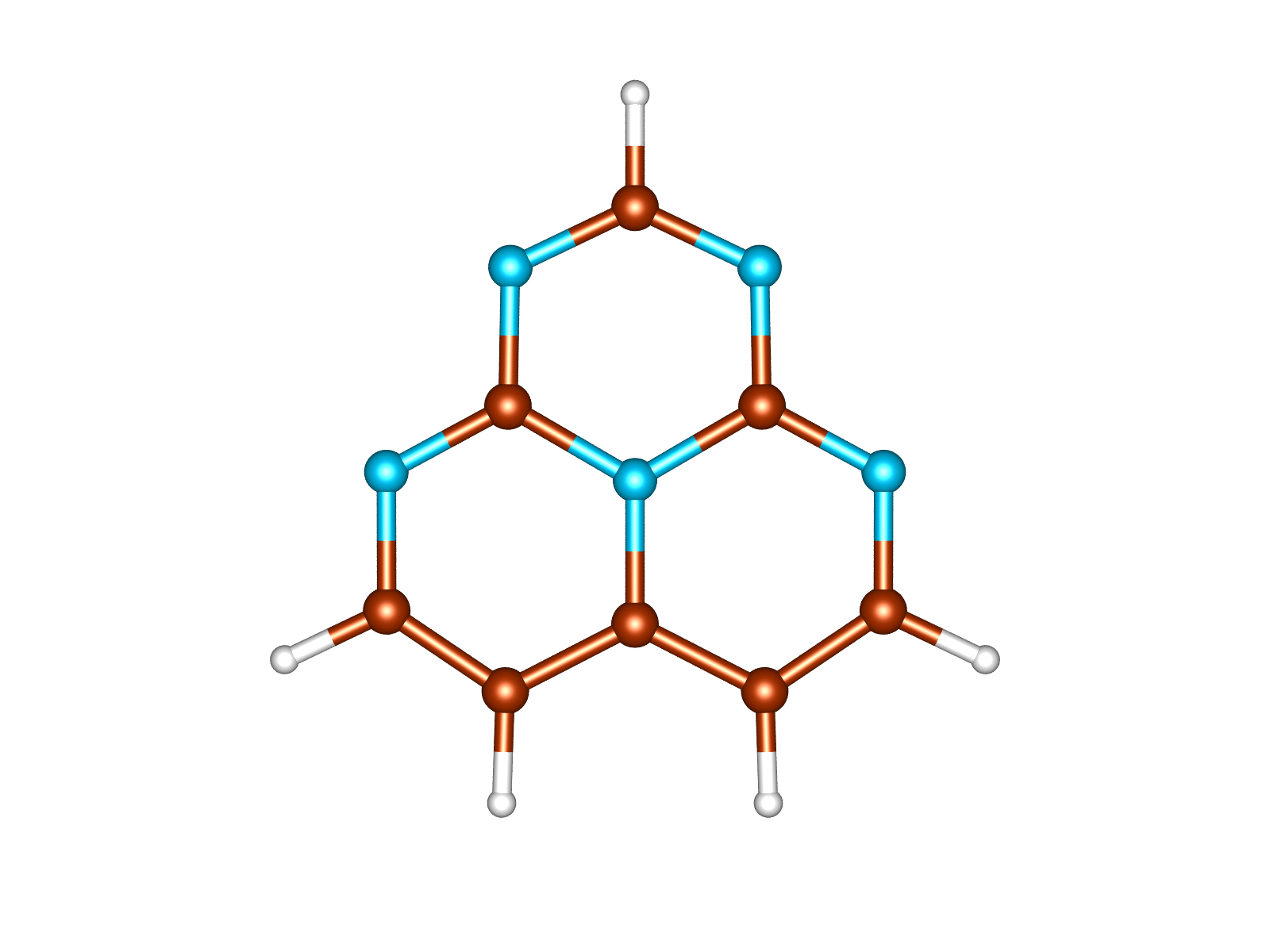}
\centering
\textbf{6}
\end{minipage}
\caption{Structures of the studied molecules. The color codes are as follows: N (blue), C (brown), and H (white).}
\label{molecules}
\end{figure}

Our analysis is focused on six heptazine-based systems, see Fig.~\ref{molecules}. Geometries were optimized at the MP2 level using the Molpro\cite{molpro} program. All results were obtained using the def2-TZVP\cite{DZ_TZ} basis set.

Complete active space self-consistent field (CASSCF) and NEVPT2 results were obtained with the Molpro~\cite{molpro} program.  We used the same software for restricted Hartree-Fock (HF), CASSCF, and Kohn-Sham DFT calculations to obtain electron integrals and one- and two-electron reduced density matrices needed for spin polarization models and adiabatic connection, AC0 and ACn, calculations. The latter were performed with the GammCor program.~\cite{gammcor}  Multireference coupled-cluster Mk-MRCCSD(T) \cite{mrcc} energies for S$_1$ and T$_1$ states computed using the CASSCF(2,2) natural orbitals were obtained with the NWChem program\cite{nwchem} and set as benchmark. CC2 results in def2-TZVP were taken from Ref.~\citenum{sundholm} while EOM-CCSD ST gaps were calculated with the Orca code.~\cite{orca}

The ground state geometry of each molecule was used for both the S$_1$ and T$_1$ states.  Except for system 4, calculations were performed  employing the \(C_{2v}\) point group symmetry, which allowed for state-specific CASSCF calculations for S$_1$ and T$_1$ states. System 4 is in a lower, \(C_s \), point group symmetry, so that two-state state-average CASSCF calculation is required to access the S$_1$ state.

Additional calculations were performed on extended test set including heptazine-derivatives obtained by substituting  C-H groups with N atoms in all possible ways. Geometries of all molecules from the extended set were optimized at B3LYP/cc-pVDZ level with no symmetry constraints using Orca code.~\cite{orca}

\section{Inverting  S$_1-$T$_1$ energy gaps by including dynamic spin polarization}

Throughout the text $\Delta E_{\rm ST}$ will denote the singlet-triplet energy gap defined as a difference of T$_1$ energy subtracted from the energy of S$_1$ state, namely
\begin{equation}
\Delta E_{\rm ST} = E({\rm S_1})-E({\rm T_1}) \ \ \ .
\end{equation}
Gap inversion would be equivalent to obtaining a negatively-valued $\Delta E_{\rm ST}$ energy difference.

In the first approximation, the S$_{1}$ and T$_{1}$ states of the considered systems can be described by singly-excited determinants, where one electron from the HOMO ($H$) orbital is excited to the orbital LUMO ($L$)
\begin{align}
\Psi_{S}^{0} &  =\frac{1}{\sqrt{2}}\left(  \left\vert \bar{H}L\right\vert
-\left\vert H\bar{L}\right\vert \right)  \ \ \ ,\label{S0}\\
\Psi_{T}^{0} &  =\frac{1}{\sqrt{2}}\left(  \left\vert \bar{H}L\right\vert
+\left\vert H\bar{L}\right\vert \right)  \ \ \ .\label{T0}%
\end{align}
A notation $\left\vert \bar{H}L\right\vert $ stands for an $N$-electron Slater determinant comprising $N/2-1$ doubly occupied orbitals and $H,L$ singly occupied orbitals. $\bar{H}$, $\bar{L}$ and $H$, $L$ indicate orbitals with $\alpha$ and $\beta$ spin components, respectively.  Additionally, we use $\{ pq \}$ and $\{ pqrs \}$ symbols to indicate which orbitals in open-shell determinants are singly-occupied without explicitly specifying their spin components. For example, both determinants $\left\vert \bar{H}L\right\vert $ and $\left\vert H \bar{L}\right\vert $ would be denoted as $\{ HL \}$.

As it is well known, the ST energy gap corresponding to a wavefunction including only $\{ HL\}$ singly excited determinants, Eqs.~\eqref{S0} and \eqref{T0}, 
\begin{equation}
\Delta E_{\rm ST}^{0} = \left\langle \Psi_{S}^{0}|\hat{H}|\Psi_{S}^{0}\right\rangle - \left\langle \Psi_{T}^{0}|\hat{H}|\Psi_{T}^{0}\right\rangle = 2\left\langle HL|LH \right\rangle \geq 0\ \ \ 
\label{E0ST}
\end{equation}
is determined by the magnitude of the HL exchange integral (throughout the text, two-electron integrals are written in the $\braket{12|12}$ convention) and is non-negative. Eq.~\eqref{E0ST} is a manifestation of the Hund's rule satisfied by wavefunctions $\Psi_{S}^{0}$ and $\Psi_{T}^{0}$. 

In the Hund's picture the ST gap is determined by the coupling of $H$ and $L$ electrons in the target spin state of the total wavefunction. This approach ignores the interaction of the unpaired $\alpha$ and $\beta$ electrons with the core electrons.  However, Kollmar and Staemmler\cite{staemmler1978violation} showed that including these interactions is not merely a quantitative refinement, but may explain the ST gap inversion in some systems. Their approach is based on extending the first-order wavefunction with double excited determinants $\{iaHL\}$, where core orbitals $i$ relax by excitations to  an arbitrary virtual orbital $a$. Including such determinants is referred to as `dynamic' spin polarization to distinguish if from the conventional `static' spin polarization. While  static spin polarization leads to unequal spin components of electron density, the  dynamic spin polarization does not lead to nonzero spin density.
Our goal is to quantify the effect of dynamic spin polarization 
for heptazine-type molecules and investigate if this mechanism is responsible for changing the sign of $\Delta E_{\rm ST}^0$ gaps.
We also investigate if sp-based gap inversion model of Kollmar and Staemmler can be employed for prescreenig molecules likely to act as INVEST systems.

We begin by following Ref.~\citenum{staemmler1978violation} and construct singlet functions by applying $i\rightarrow a$ excitations to determinants in the $\Psi_{S}^{0}$ function, which leads to two functions 
\begin{align}
\Psi_{S}^{1} &  =\frac{1}{\sqrt{12}} \left( 2|\bar{i}\bar{a}HL| + 2|ia\bar{H}\bar{L}| -|\bar{i}a\bar{H}L| - |\bar{i}aH\bar{L}| -|i\bar{a}\bar{H}L| - |i\bar{a}H\bar{L}| \right)  \ \ \ ,\label{S1} \\
\Psi_{S}^{1^{\prime}} &  =\frac{1}{2}(|i\bar{a}H\bar{L}|-|i\bar{a}\bar
{H}L|-|\bar{i}aH\bar{L}|+|\bar{i}a\bar{H}L|)\ \ \ .
\end{align}
Analogously, three triplet $M_{S}=0$ functions  can be generated from $\Psi_T^0$ by considering $i\rightarrow a$
excitations and they read
\begin{align}
\Psi_{T}^{1} &  =-\frac{1}{2} ( |\bar{i}a\bar{H}L| -|\bar{i}aH\bar{L}| +|i\bar{a}\bar{H}L|-|i\bar{a}H\bar{L}|)\ \ \ ,\label{T1}\\
\Psi_{T}^{2} &  =-\frac{1}{\sqrt{2}}(|\bar{i}\bar{a}HL|-|ia\bar{H}\bar
{L}|)\ \ \ ,\\
\Psi_{T}^{1^{\prime}} &  =-\frac{1}{2}(|\bar{i}a\bar{H}L|+|\bar{i
}aH\bar{L}|-|i\bar{a}\bar{H}L|-|i\bar{a}H\bar{L}|)\ \ \ .\label{T1'}%
\end{align}
Recall that from the Epstein-Nesbet perturbation theory (PT), the drop in the unperturbed energy, $E^0 = \langle \Psi^{0}|\hat{H}|\Psi^0 \rangle$, that results from including in the wavefunction a configuration $\Psi'$ and corresponds to the second-order correction, is given by $|\langle \Psi^{0}|\hat{H}|\Psi'\rangle|^2/(E^{0}-E')$, where $E'=\langle \Psi'|\hat{H}|\Psi'\rangle $. After
evaluating PT terms for functions given in Eqs.~\eqref{S1}--\eqref{T1'}, see Supporting Information for explicit expressions of Hamiltonian elements and energy differences in
terms of two-electron integrals, and subtracting triplet contributions from
the singlet ones, the following spin-free expression is obtained
\begin{equation}
\Delta E_{i,a}^{sp}=\frac{3}{2}\frac{\left(  \left\langle iH|Ha\right\rangle
-\left\langle iL|La\right\rangle \right)  ^{2}}{E_{S}^{0}-E_{S}^{1}}-\frac
{1}{2}\frac{\left(  \left\langle iH|Ha\right\rangle -\left\langle
iL|La\right\rangle \right)  ^{2}}{E_{T}^{0}-E_{T}^{1}}-\frac{\left(
\left\langle iH|Ha\right\rangle +\left\langle iL|La\right\rangle \right)
^{2}}{E_{T}^{0}-E_{T}^{2}}\ \ \ \label{Espia}%
\end{equation}
in agreement with Ref.~\citenum{staemmler1978violation} (notice that contributions to $\Delta E_{i,a}^{sp}$ from states $\Psi
_{S}^{1^{\prime}}$ and $\Psi_{T}^{1^{\prime}}$, not considered in Ref.\citenum{staemmler1978violation}, cancel
each other).  $\Delta E_{i,a}^{sp}$ represents a contribution to the ST\ gap
from spin polarization of two electrons occupying the orbital $i$ by allowing their excitation to the orbital $a$. To ease its interpretation, expression in Eq.~\eqref{Espia} can be approximated by assuming a common denominator given as Hartree-Fock orbital energy difference
$\varepsilon_{i}-\varepsilon_{a}$ for all three terms, leading to
\begin{equation}
\Delta E_{i,a}^{sp}\approx4\frac{\left\langle iH|Ha\right\rangle \left\langle
iL|La\right\rangle }{\varepsilon_{a}-\varepsilon_{i}}\ \ \ .\label{spia}
\end{equation}
It is now clear that spin polarization energy, $\Delta E_{i,a}^{sp}$, pertaining to a pair of orbitals $(i,a)$, where $i<H$ and $a>L$, stabilizes the singlet state with respect to triplet if orbitals $i$ and $a$ overlap significantly with both $H$ and $L$ orbitals and, taking into account that $\epsilon_a-\epsilon_i > 0 $, exchange integrals $\braket{iH|Ha}$ and $\braket{iL|La}$ are of the opposite sign. The magnitude of the $\Delta E_{i,a}^{sp}$ energy is dependent on the closeness of the $i$ and $a$ orbital energy levels.

For the studied systems the $\left(H-1,H-2\right)$ and  $\left(L+1,L+2\right)$ pairs of orbitals are
degenerate, so the simplest model for the ST gap accounting for major spin polarization would include two pairs of orbitals, and the ST energy gap expression of such a ``12'' model would read
\begin{equation}
\Delta E_{\rm ST}^{sp_{12}} = \Delta E_{\rm ST}^{0} + \Delta E_{H-1,L+1}^{sp} + \Delta E_{H-2,L+2}^{sp}\ \ \ \ .
\label{STsp12}
\end{equation}
For system 4, the model included also the contributions from the other two combinations of orbitals $\Delta E_{H-1,L+2}^{sp} + \Delta E_{H-2,L+1}^{sp}$, which are negligible in the other systems due to point group symmetry.
To fully account for spin polarization, considered pairs of orbitals $(i<H,a>L)$  must fulfill conditions leading to ST gap reversal, formulated below Eq.~\eqref{spia}: $i$) orbitals within each pair should strongly overlap with $H$ and $L$ orbitals,  $ii$) the exchange integrals $\braket{iH|Ha}$ and $\braket{iL|La}$ are of the opposite signs.
For heptazine derivatives the $\pi$ orbitals satisfy  these requirements and the all-$\pi$ sp-inclusive ST energy gap reads
\begin{equation}
\Delta E_{\rm ST}^{sp_{\pi}} = \Delta E_{\rm ST}^{0} + \sum_{i<H}\sum_{a>L}\Delta E_{i,a}^{sp} \ \ \ ,\label{STall}
\end{equation}
(orbitals $i$ and $a$ are of the $\pi$-type).

\begin{figure}
\centering
\begin{minipage}[b]{0.32\textwidth}
\includegraphics[scale=0.1]{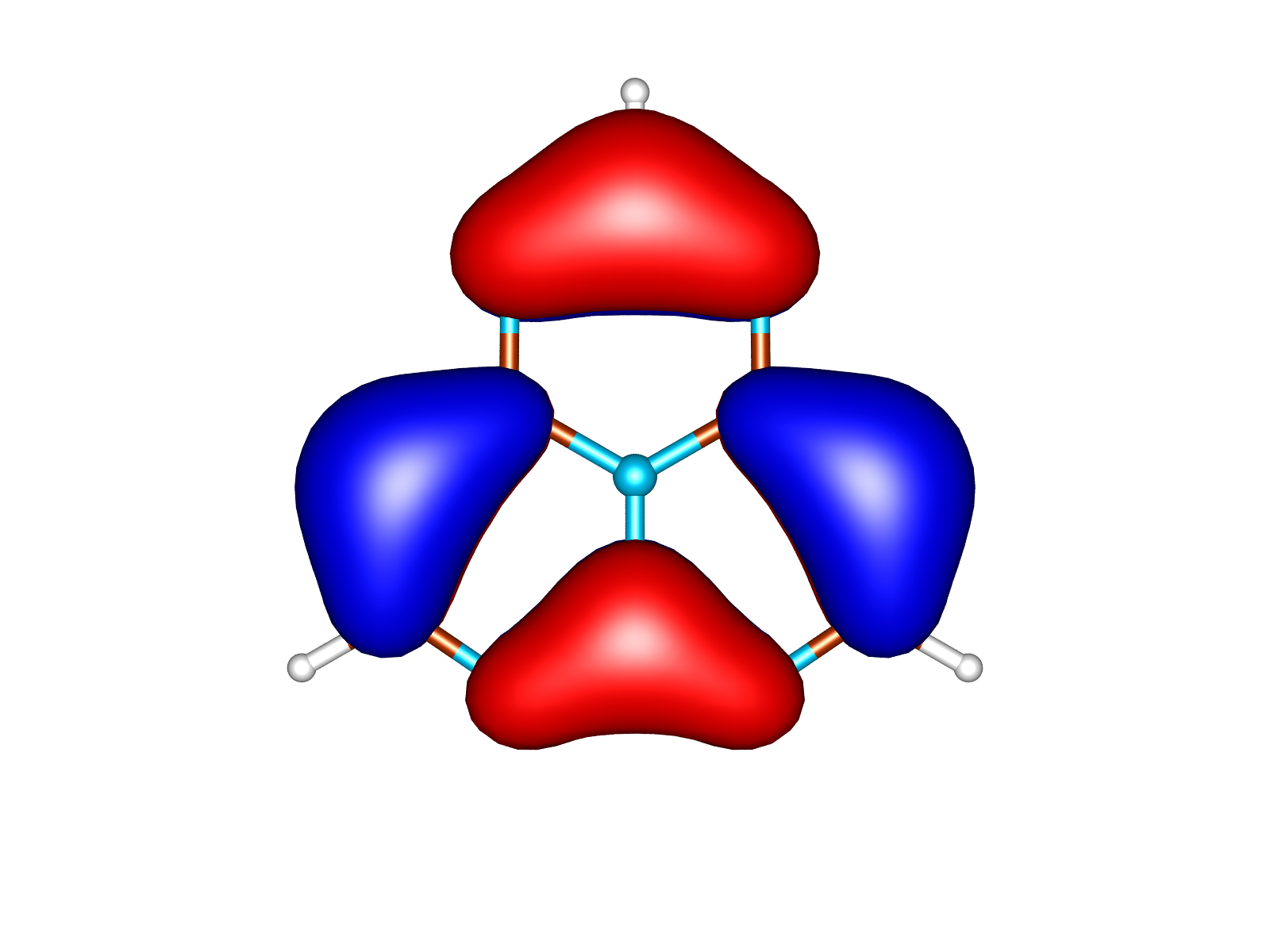}
\text{HOMO--2}
\end{minipage}
\begin{minipage}[b]{0.32\textwidth}
\includegraphics[scale=0.1]{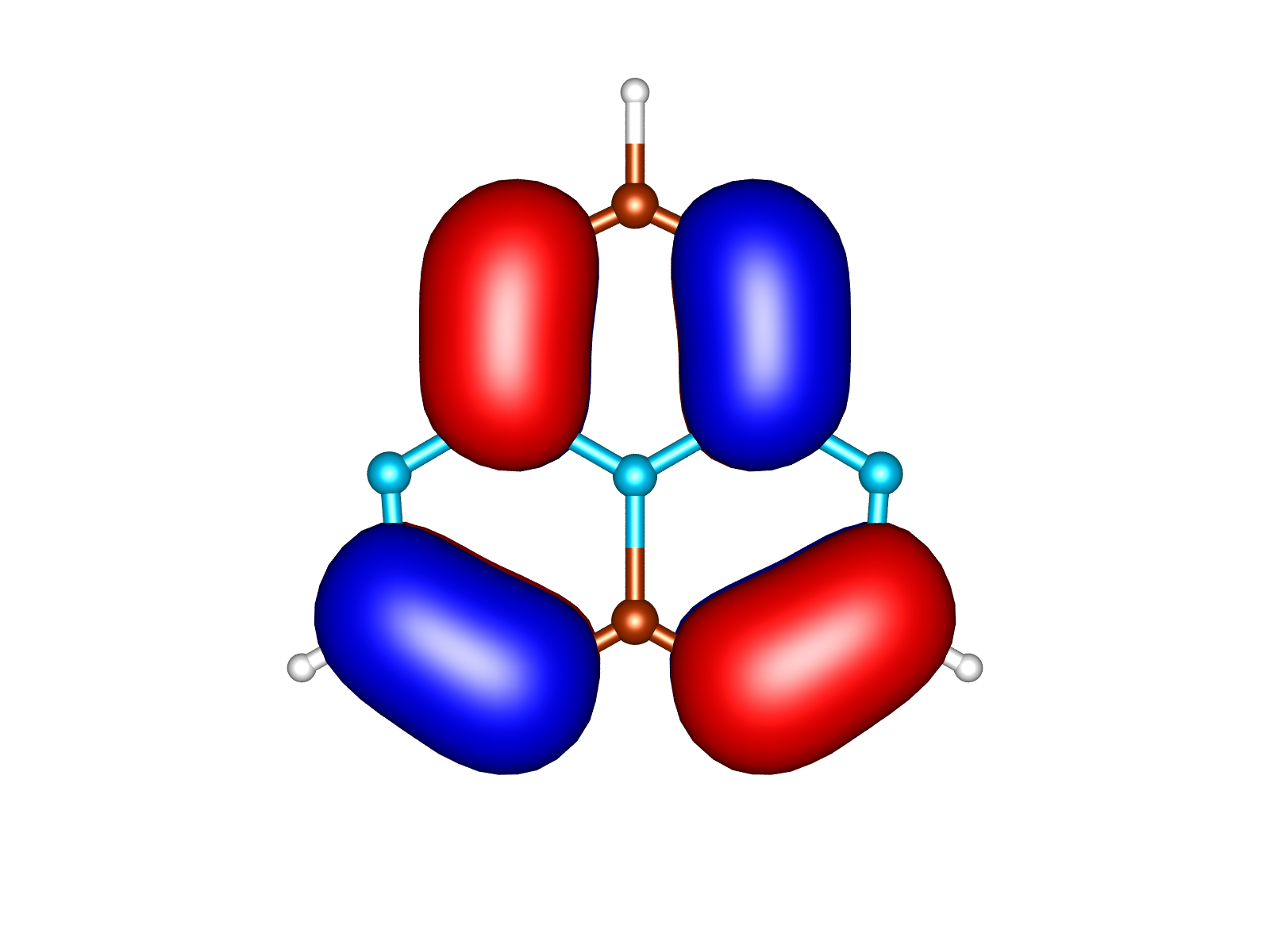}
\text{HOMO--1}
\end{minipage}
\begin{minipage}[b]{0.32\textwidth}
\includegraphics[scale=0.1]{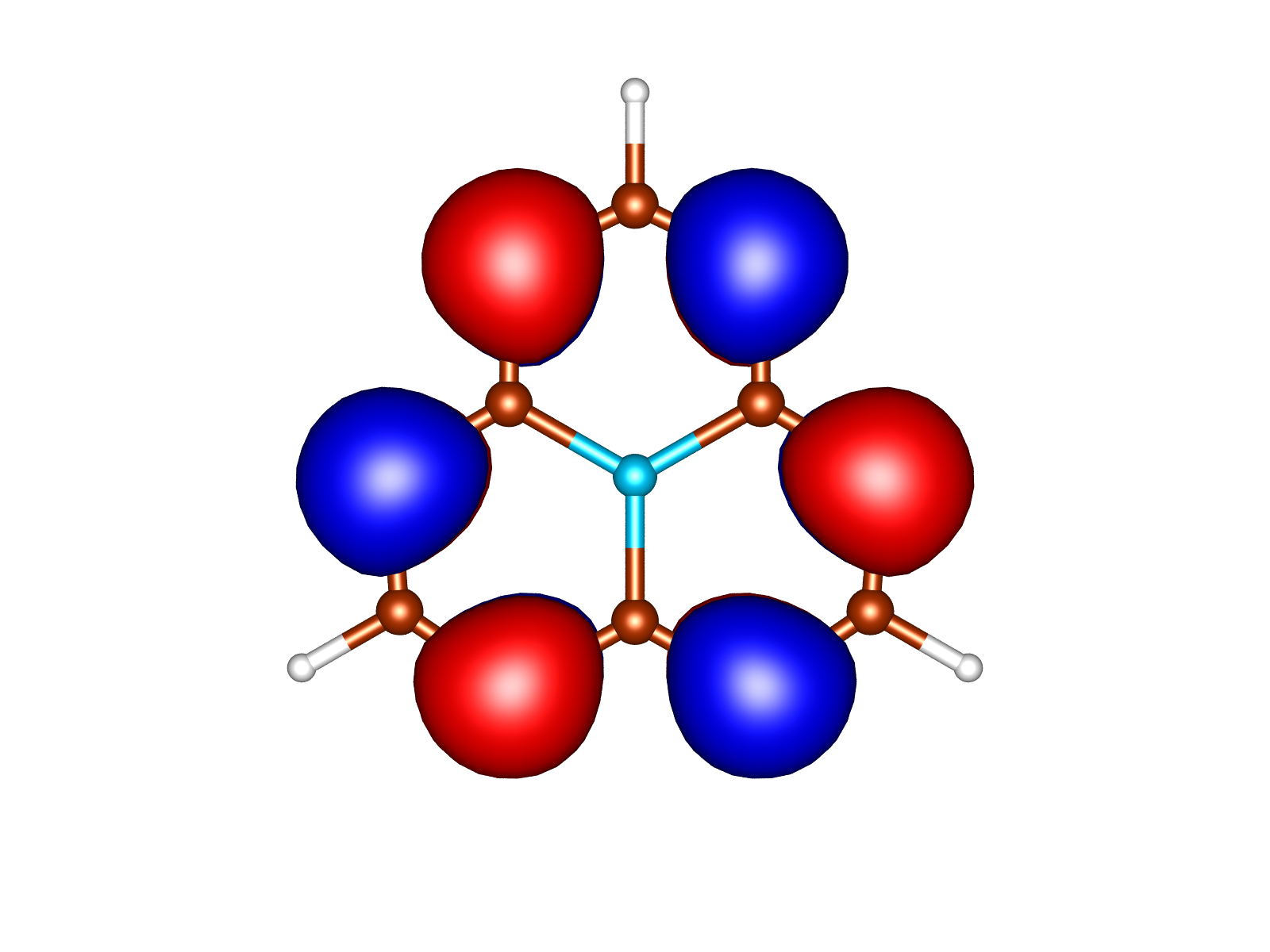}
\text{HOMO}
\end{minipage}
\begin{minipage}[b]{0.32\textwidth}
\includegraphics[scale=0.1]{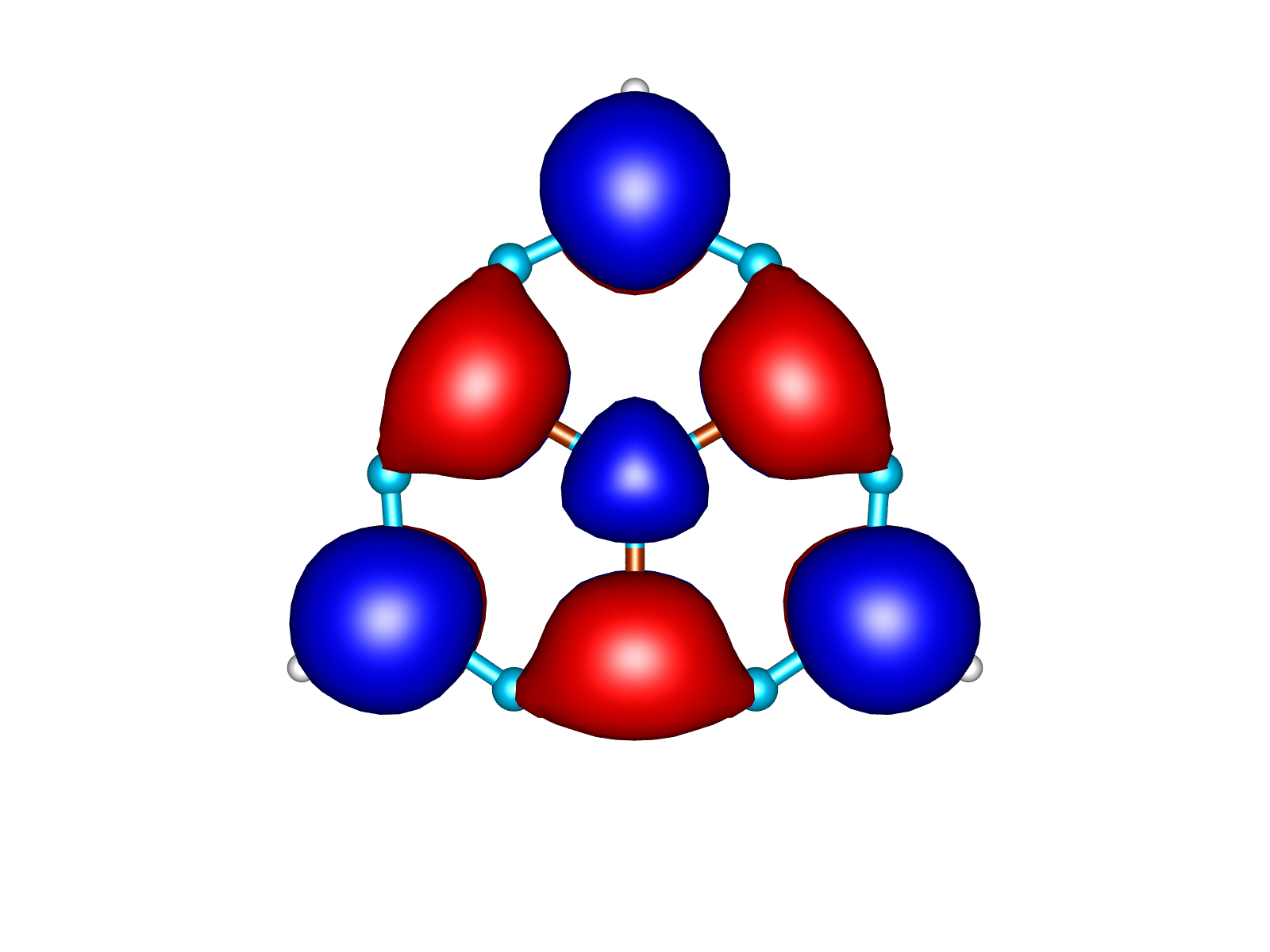}
\text{LUMO}
\end{minipage}\begin{minipage}[b]{0.32\textwidth}
\includegraphics[scale=0.1]{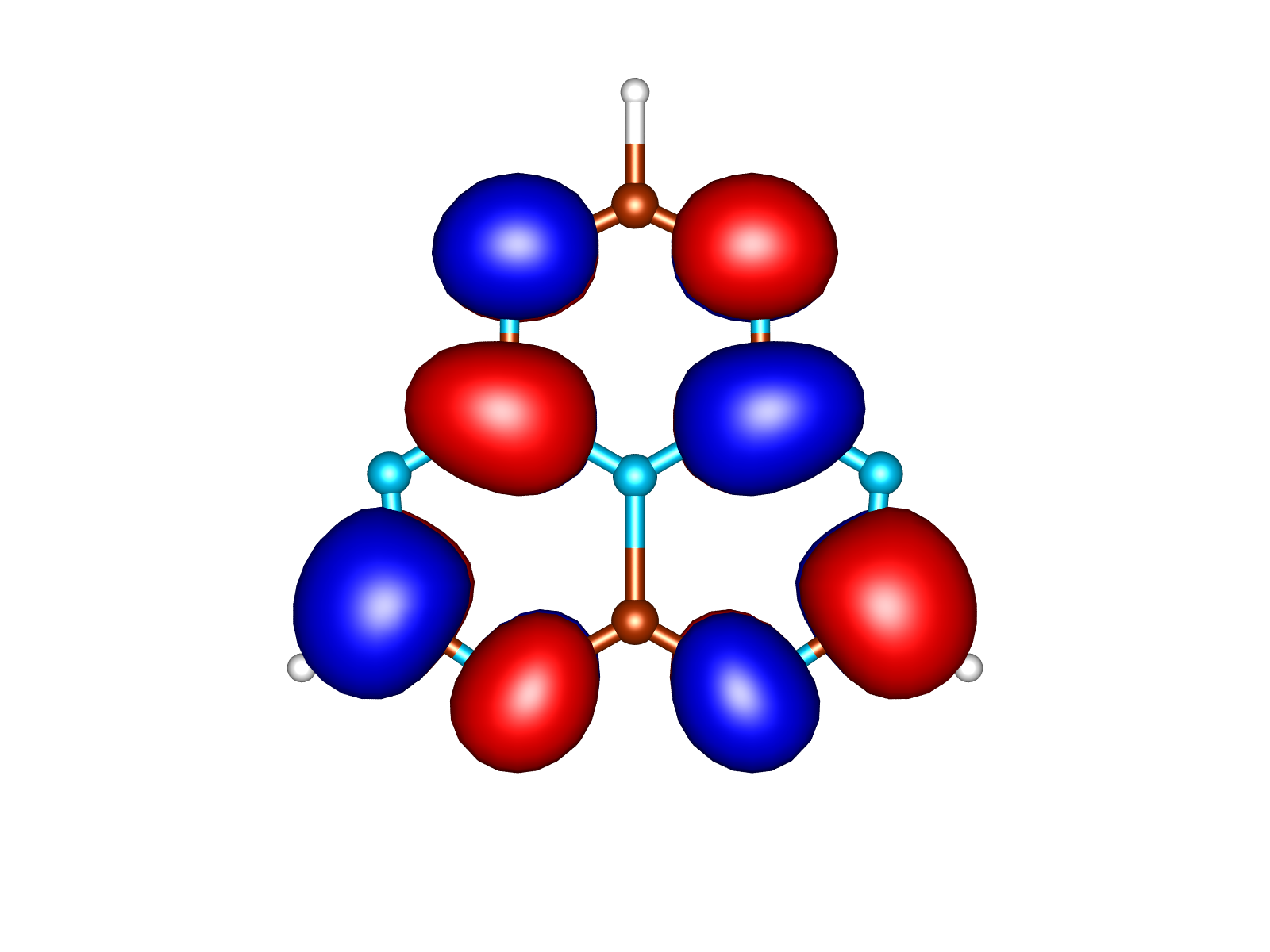}
\text{LUMO+1}
\end{minipage}\begin{minipage}[b]{0.32\textwidth}
\includegraphics[scale=0.1]{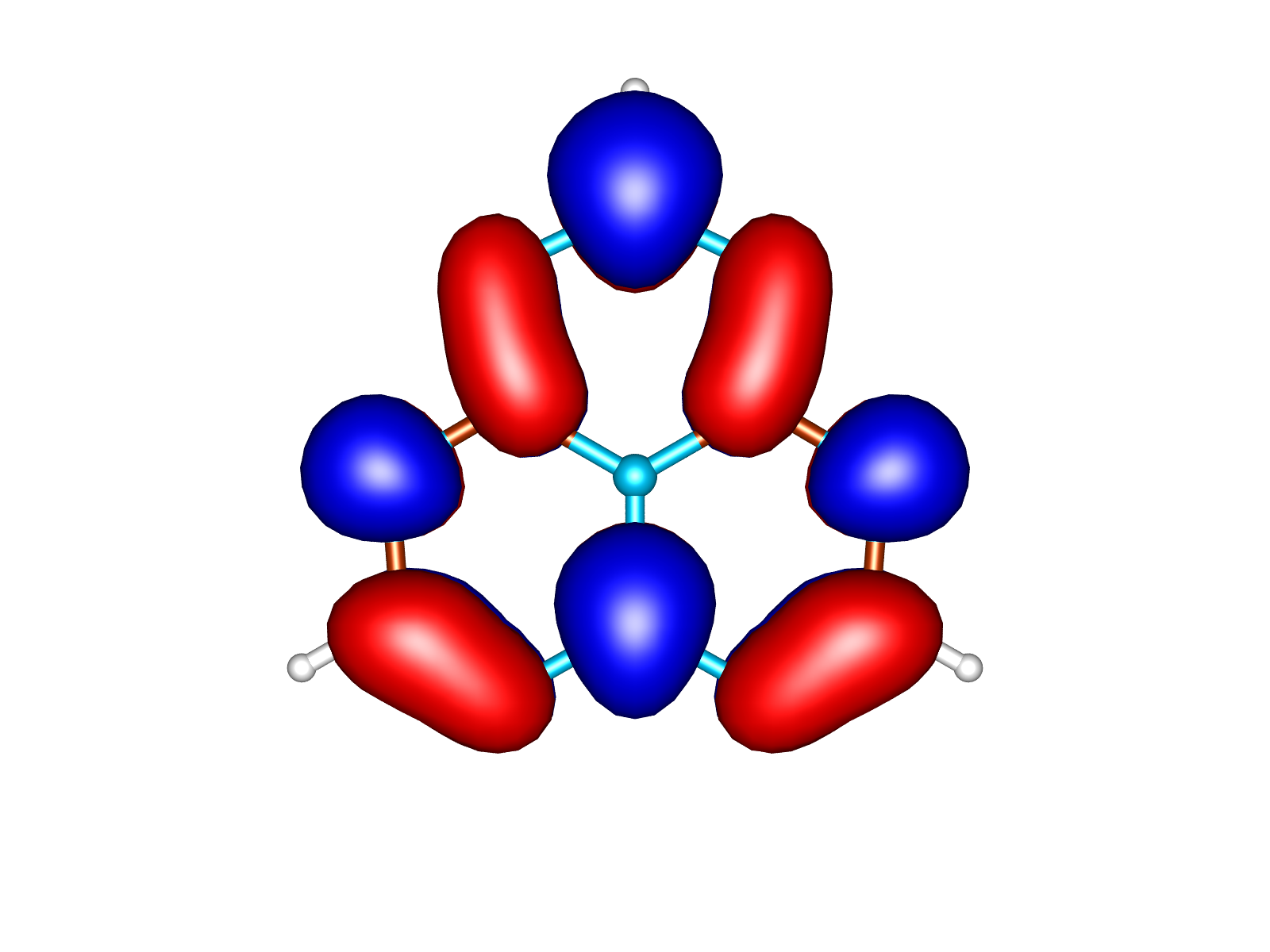}
\text{LUMO+2}
\end{minipage}
\caption{Natural orbitals of system 2 corresponding to state-averaged CASSCF(14,14) calculations with one singlet and one triplet states.}
\label{hepta_orbs_sing6}
\end{figure}

We consider two sets of orbitals in further analysis. The first set is given by canonical HF orbitals. The second one corresponds to a ground state CASSCF(14,14) wavefunction with the active space  containing all the $\pi$ electrons and orbitals (2p$_{\text{z}}$ orbitals on carbon and nitrogen atoms plus one virtual orbital of the ``double-shell'' 3p$_{\text{z}}$ character\cite{sundholm}). The CASSCF(14,14) natural occupation numbers for systems 2 and 4, used as illustrative cases, are given in Table~\ref{TAB:occupations}. Notice that orbitals $H$ and $L$ pertain to $(N/2)$th and $(N/2+1)$th orbitals, respectively,  of the occupation numbers close to $0.5$. These orbitals are depicted in Figure~\ref{hepta_orbs_sing6} (see also Figures~1-12 in Supporting Information) together with four other most strongly correlated, i.e.\ of the occupancies deviating most from $0$ and $1$, orbitals.

\begin{table}[]
\caption{Natural occupation numbers for systems 2 and 4 from state-averaged CASSCF(14,14) calculations, for system 2 corresponding to the natural orbitals presented in Figure \ref{hepta_orbs_sing6}. }

\renewcommand{\arraystretch}{1.25}
\begin{tabular}{c c c c c}
         & \multicolumn{2}{c}{Sys 2} & \multicolumn{2}{c}{Sys 4} \\
         \hline
Orbital  & S$_1$           & T$_1$            & S$_1$            & T$_1$            \\
\hline
$H$ - 6 & 0.995     & 0.993     & 0.990     & 0.989     \\
$H$ - 5 & 0.969     & 0.970     & 0.971     & 0.971     \\
$H$ - 4 & 0.967     & 0.968     & 0.970     & 0.971     \\
$H$ - 3 & 0.967     & 0.968     & 0.969     & 0.971     \\
$H$ - 2 & 0.928     & 0.933     & 0.936     & 0.939     \\
$H$ - 1 & 0.929     & 0.933     & 0.933     & 0.936     \\
$H$     & 0.513     & 0.504     & 0.639     & 0.509     \\
$L$     & 0.485     & 0.492     & 0.359     & 0.486     \\
$L$ + 1 & 0.074     & 0.073     & 0.070     & 0.071     \\
$L$ + 2 & 0.074     & 0.073     & 0.069     & 0.070     \\
$L$ + 3 & 0.036     & 0.035     & 0.033     & 0.031     \\
$L$ + 4 & 0.036     & 0.035     & 0.032     & 0.031     \\
$L$ + 5 & 0.026     & 0.023     & 0.024     & 0.022     \\
$L$ + 6 & 0.001     & 0.001     & 0.004     & 0.004    \\
\hline
\multicolumn{5}{c}{CI coefficients} \\
\hline
$\{ HL \}$ & 68.6\% & 69.3\% & 67.3\% & 71.3\% \\
$\{(H-p) \ (L+p) \ H \ L \}_{p=1,2}$ & 8.2\% & 7.8\% & 7.2\% & 0.3\% \\
$\{ i \ a\ H \ L\}_{i,a\in \pi}$
& 15.0\% & 14.6\% & 13.4\% & 5.6\% \\ \hline\hline
\end{tabular}
\label{TAB:occupations}
\end{table}

The results for ST gaps obtained without spin polarization, $\Delta E^0_{\rm ST}$, with partial and full spin polarization, $\Delta E_{\rm ST}^{sp_{12}}$ and $\Delta E_{\rm ST}^{sp_{\pi}}$ models, respectively, are presented in Table \ref{TAB:HF_CASSCF}. 
First, it can be seen that $\Delta E^0_{\rm ST}$ gaps are positive and small, which indicates that for all considered systems the $H$ and $L$ orbitals overlap marginally both at the HF and CASSCF(14,14) levels of theory. Accounting for dynamic spin polarization from only two pairs of orbitals, $(H-1,L+1)$ and $(H-2,L+2)$, see Eq.~\eqref{STsp12}, significantly reduces ST gaps. Using canonical HF\ orbitals leads to negative ST gaps for systems 1 and 2, while with CASSCF(14,14)\ natural orbitals all ST gaps turn negative. Considering all $(i<H,a>L)$ pairs of $\pi$ orbitals, Eq.~\eqref{STall}, 
removes another $0.2-0.3$~eV from the ST gap, regardless of the employed orbitals. We conclude that the PT-based model for dynamic spin polarization substantially lowers the S$_1$ state energy with respect to T$_1$, possibly leading to sign reversal of the ST gap. The combined sp effect of two pairs of orbitals, $(H-1,L+1)$ and $(H-2,L+2)$, accounts for more than half of the ST gap reduction.

\begin{table}
\caption{S$_1-$T$_1$ energy gaps obtained without spin polarization, $\Delta E_{\rm ST}^{0}$,
with spin polarization computed
from two pairs of $\pi$ orbitals, 
$\Delta E_{\rm ST}^{{\rm sp}_{12}}$,
from all occupied-virtual pairs of $\pi$ orbitals, $\Delta E_{\rm ST}^{{\rm sp}_{\pi}}$,
and from CI with $\{H \ L\}$, $\{H-1 \ L+1 \ H \ L \}$, and $\{H-2 \ L+2 \ H \ L \}$ determinants, $\Delta E_{\rm ST}^{{\rm CI}_{12}}$. 
HF canonical and CASSCF(14,14) orbitals were used. The latter followed from state-averaged CASSCF calculations with one singlet and one triplet states. All values in eV.}
\renewcommand{\arraystretch}{1.5} 
\begin{tabular}{c c S S S S S S} \hline
\multicolumn{1}{l}{Orbitals}  &  ST gap & 1 & 2 & 3 & 4 & 5 & 6 \\ \hline
\multirow{4}{*}{HF} & $\Delta E_{\rm ST}^{0} $                 & 0.25  & 0.25  & 0.49  & 0.67  & 0.68  & 0.61  \\
 & $\Delta E_{\rm ST}^{sp_{12}}$      & -0.16 & -0.18 &  0.09 & 0.29 & 0.33 & 0.22  \\
 & $\Delta E_{\rm ST}^{sp_{\pi}} $    & -0.37 & -0.45 & -0.11 & 0.09 & 0.07 & 0.00  \\
 & $\Delta E_{\rm ST}^{\rm CI_{12}} $     & -0.21 & -0.25 &  0.10 & 0.27 & 0.10 & 0.08  \\ \hline
\multirow{4}{*}{CASSCF} 
  & $\Delta E_{\rm ST}^{0} $        & 0.19  & 0.19  & 0.25  & 0.32  & 0.25  & 0.28  \\                   
  & $\Delta E_{\rm ST}^{sp_{12}}$   & -0.33 & -0.40 & -0.28 & -0.19 & -0.31 & -0.27 \\
  & $\Delta E_{\rm ST}^{sp_{\pi}}$  & -0.60 & -0.73 & -0.54 & -0.46 & -0.61 & -0.55 \\
  & $\Delta E_{\rm ST}^{\rm CI_{12}}$ & -0.29 & -0.36 & -0.25 &  -0.14 & -0.28 & -0.24 \\ \hline
\end{tabular}
\label{TAB:HF_CASSCF}
\end{table}

Spin polarization in the wavefunction picture is manifested by the presence of doubly-excited determinants. To check if the simple PT-based model for sp yields qualitatively correct predictions, we have performed CI calculations constructing the S$_1$ and T$_1$ wavefunctions  from $\{H \ L\}$, $\{H-1 \ L+1 \ H \ L \}$, and $\{H-2 \ L+2 \ H \ L \}$ singly- and doubly-excited determinants, employing either HF or CASSCF(14,14) orbitals. The CI results, 
denoted as $\Delta E_{\rm ST}^{\rm CI_{12}}$ in Table~\ref{TAB:HF_CASSCF}, are in close agreement with the $\Delta E_{\rm ST}^{sp_{12}}$ model. 
This suggests that  PT-based approach can be used as a cost-saving alternative to CI calculations in prescreening for molecules with inverted gaps.

ST gaps obtained from $\Delta E_{\rm ST}^{sp_{12}}$ and $\Delta E_{\rm ST}^{sp_{\pi}}$  should be considered as basic approximations due to the fact that there are only
$\{HL\}$ and $\{iaHL \}$ determinants in the wavefunctions. Thus, the model gaps cannot be quantitatively correct.
A comparison with Mk-MRCCSD(T) results, shown in the last column of Table~\ref{TAB:ST}, reveals  that ST gaps predicted by sp$_{12}$ and sp$_\pi$ models employed with HF orbitals, cf.\ Table \ref{TAB:HF_CASSCF}, deviate on average by 0.26 and 0.15 eV from the Mk-MRCCSD(T) benchmarks, respectively. With CASSCF(14,14) orbitals the respective mean absolute errors amount to 0.14 and 0.43~eV. Such inaccuracies result from the lack of dynamic electron correlation in the considered models. This problem is addressed in the next section.

\begin{figure}
\centering
\centering
\begin{minipage}[b]{0.98\textwidth}
\includegraphics[scale=0.30]{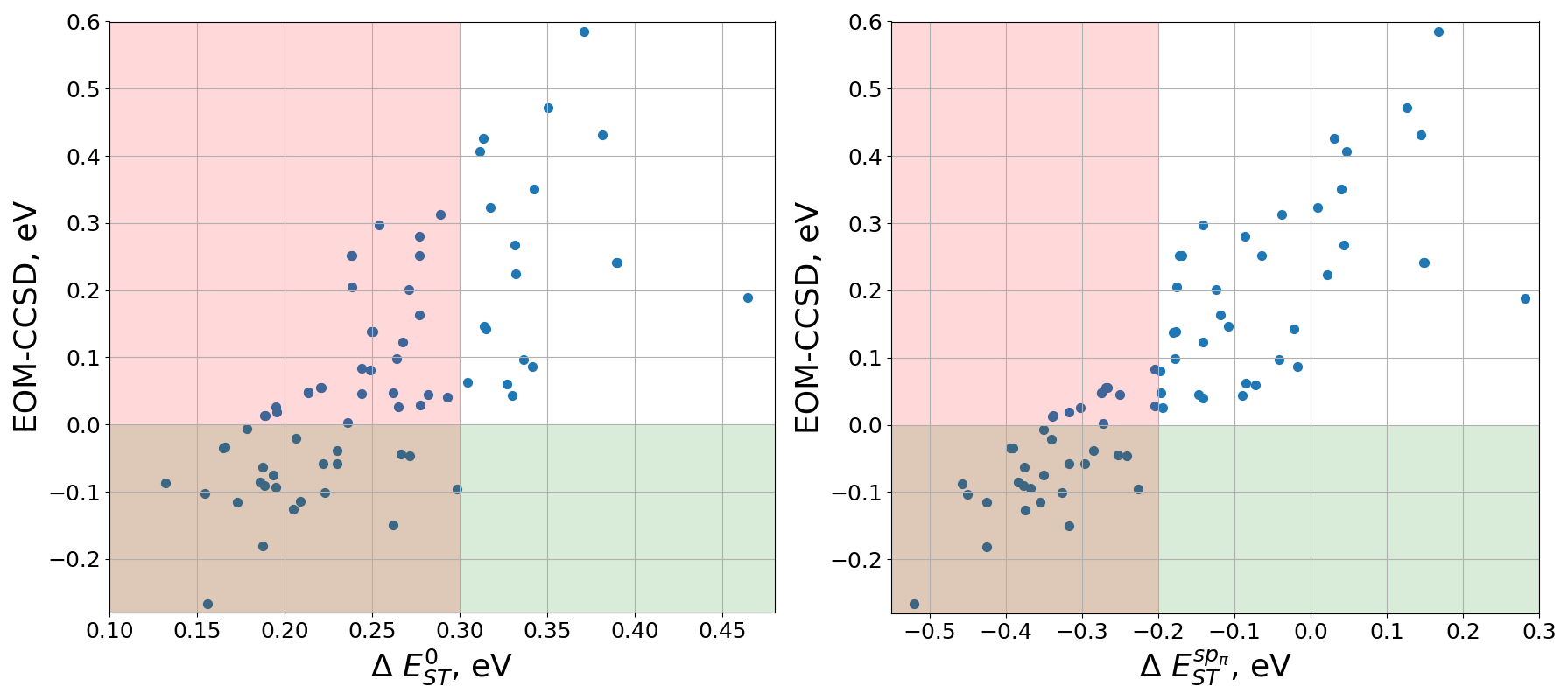}
\centering
\end{minipage}
\caption{Left panel: ST energy gaps without spin polarization, Eq.~\eqref{E0ST}. Right panel:
ST energy gaps with spin polarization from all occupied-virtual $\pi$ orbital pairs, Eqs.~\ref{STall} and \eqref{Espia}. Results obtained with BLYP ground state orbitals vs. EOM-CCSD values for 
extended test set of molecules.}
\label{PLOT:DFT}
\end{figure}

To further explore if the sp effect is decisive in inverting the ST\ gap and if the model presented in Eq.~\eqref{STall} would be useful in screening for INVEST candidates, we have applied it to an extended test set of molecules including heptazine-derived systems obtained by substitution of C-H with N in all possible ways, similar as in Ref.~\citenum{guzik2021organic}. The ST gaps have been computed using Eqs.~\eqref{Espia} and \eqref{STall}. Initial tests on systems 1-6 have shown, see Table \ref{TAB:HF_CASSCF}, that CASSCF(14,14) orbitals satisfy conditions required to invert the ST gaps via the sp mechanism to a greater extent than HF orbitals and, unlike the latter, they have led to obtaining negative gaps for all molecules. This suggests that correlated orbitals are a better choice for INVEST screening. Guided by this finding, we have used KS-DFT orbitals from ground-state BLYP\cite{b88,lyp1,lyp2} calculations for tests on our extended set of molecules. In Figure~\ref{PLOT:DFT}, 
we present ST energy gaps obtained without spin polarization, see Eq.~\eqref{E0ST}, and with spin polarization fully accounted for, cf.\ Eq.~\eqref{STall}, compared with the EOM-CCSD values. Evidently, employing only the HOMO-LUMO exchange integral for INVEST\ prescreening does a poor job, as hardly any correlation between $\Delta E_{\rm ST}^{0}$ and EOM-CCSD ST gap values can be seen. Shifting the $\Delta E_{\rm ST}^{0}$ values by $-0.30$ eV and keeping only molecules with resulting negative gaps would leave us with too large set: not only molecules with EOM-CCSD-predicted negative gaps would be included, but also most of those with positive gaps, see left panel in Figure ~\ref{PLOT:DFT}. Thus, a criterion based on $\Delta E_{\rm ST}^{0}$ values is not sufficiently selective in screening for INVEST systems. A satisfactory correlation is obtained by employing the full sp$_\pi$ model with KS-DFT orbitals and shifting the $\Delta E_{ST}^{sp_\pi}$ energies by $+0.2$~eV. This value is recommended in future INVEST prescreening calculations, as it results in a good overlap between the subsets of molecules with positive and negative gaps obtained from the sp model and EOM-CCSD, as shown in Figure ~\ref{PLOT:DFT}, right panel. Notice that the correlation is worse with HF orbitals, see the right panel of Figure~13 in Supporting Information.

\begin{figure}
\centering
\centering
\begin{minipage}[b]{0.98\textwidth}
\includegraphics[scale=0.3]{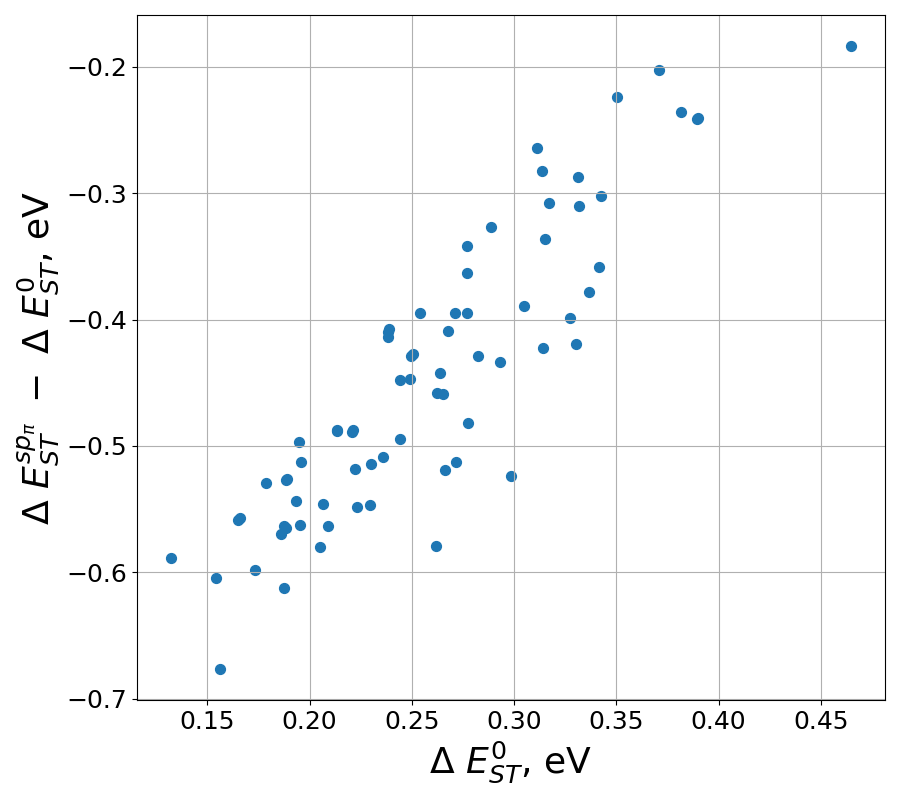}
\includegraphics[scale=0.32]{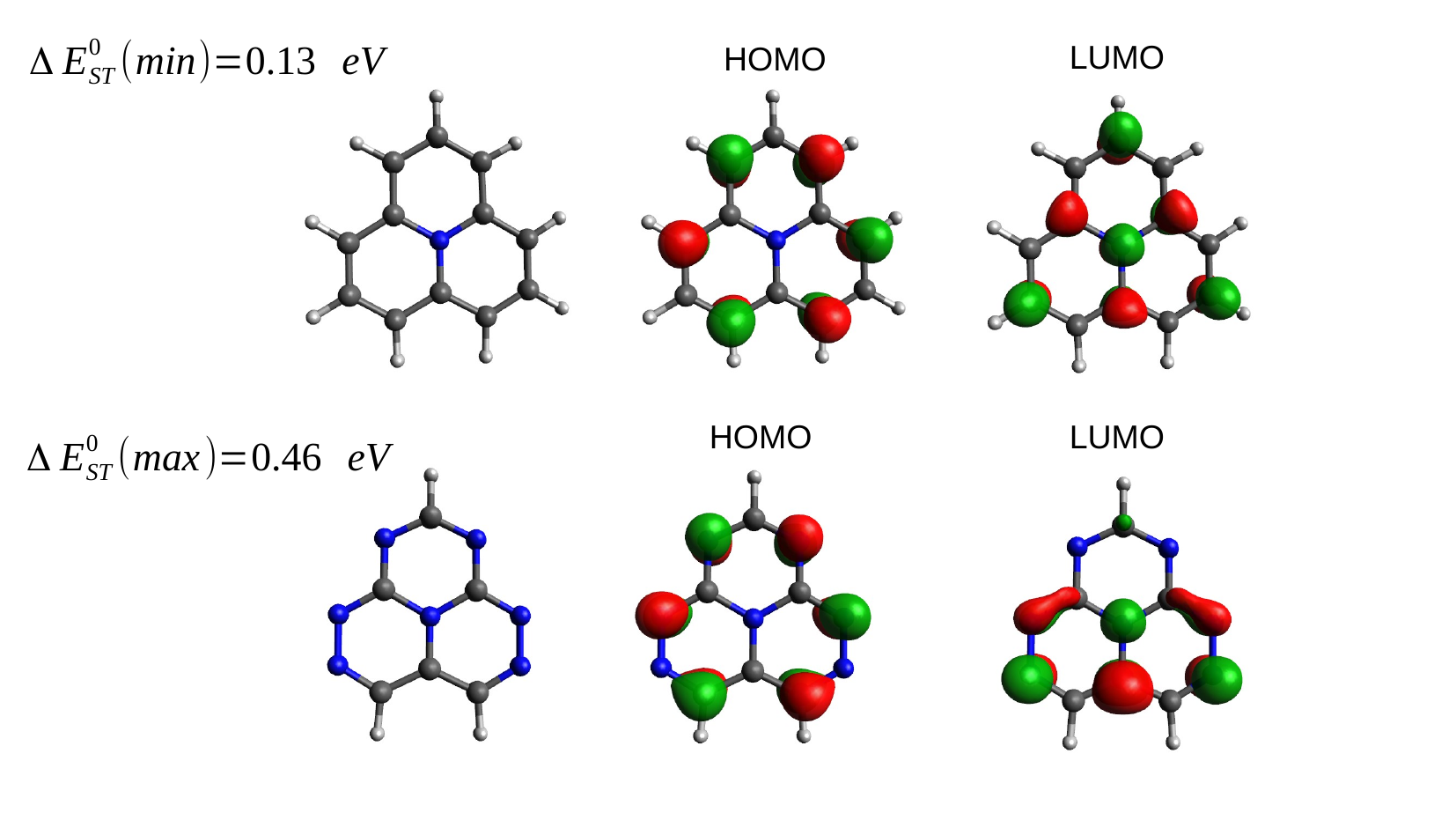}
\centering
\end{minipage}
\caption{Left panel: Contributions from spin polarization to ST energy gaps, second term of the right-hand-side of Eq.~\eqref{STall}, vs.\ ST energy gaps without spin polarization, Eq.~\eqref{E0ST}. 
Right panel: Molecules corresponding to the lowest and highest H-L exchange interaction. Results obtained with BLYP ground state orbitals.}
\label{PLOT:DFT2}
\end{figure}

Another observation following from calculations on the extended set of molecules is that the sp effect is approximately inversely proportional to the HL exchange interaction, see  Figure \ref{PLOT:DFT2}. Therefore, larger HL integrals correspond to lower spin polarization, both effects being detrimental to obtaining negative ST gaps. 
Apparently, sp for the considered isoelectronic molecules is as sensitive as HL exchange integrals to the composition and distribution of carbon and nitrogen atoms in heptazine derivatives.

In Ref.\citenum{ricci2022establishing} low HL exchange interaction has been associated with high symmetry point groups $D/C_{3h}$, $C_{3v}$, $C_{2v}$. In particular, the existence of the $\sigma_v$ plane was identified as critical in minimizing the HL overlap. Since, as we have shown,  minimization of the latter is accompanied by maximization of the sp effect, in general it seems to be a good strategy to account for symmetry while designing INVEST molecules. However, symmetry cannot be used as the only criterion in high-throughput screening. Recall, that system 4 does not possess $\sigma_v$ plane as symmetry element (Figure \ref{molecules}), but the corresponding ST gap is negative. In contrast, the molecule with the largest HL exchange integral and most positive $\Delta E_{ST}^{sp_\pi}$ gap, cf.\ Figure \ref{PLOT:DFT2}, belongs to the $C_{2v}$ point group.

Results obtained on the extended set show that a simple spin polarization model considered in this section used with KS-DFT orbitals can serve as a computationally inexpensive predictor of INVEST\ molecules. 
Notice that for systems 1-6 a common-denominator approximation adopted in Eq.~\eqref{spia} yields energy gaps
in a good agreement with those obtained if sp from all $\pi$ orbital-pairs is included as in Eq.~\eqref{Espia}, see Table~1 in Supporting Information. Thus, further simplification of the model is achievable if Eq.~\eqref{spia} is used instead of Eq.~\eqref{Espia}. Compared to predictors used in Ref.~\citenum{guzik2021organic} based on a double-hybrid DFT functional the proposed model is computationally more efficient (it requires performing a ground state calculation with a semi-local functional) and above all physically meaningful, as it explicitly includes the sp effect responsible for the gap inversion.

\section{The effect of dynamic correlation energy}

CAS wavefunctions should effectively capture the sp effect. One expects that expanding the active space from (2,2), two electrons on $H$ and $L$ orbitals, via (6,6), six electrons on $H$, $L$, and two pairs of degenerate orbitals $H-1$, $H-2$ and $L+1$, $L+2$, up to (14,14), all $\pi$ electrons on all $\pi$ orbitals, would lead to ST energy gaps  that reflect the $\Delta E^0_{ST} > \Delta E^{sp_{12}}_{\rm ST} > \Delta E^{sp_\pi}_{\rm ST}$ relation observed for PT-based models, namely:
\begin{equation}
\Delta E_{\rm ST}^{\rm CASSCF(2,2)}> \Delta E_{\rm ST}^{\rm CASSCF(6,6)} > \Delta E_{\rm ST}^{\rm CASSCF(14,14)} \ \ \ .\label{ineq}
\end{equation}
Indeed, CASSCF energy gaps presented in Table~\ref{TAB:ST} agree with the predicted inequality relations, Eq.~\eqref{ineq}, for each system (see the column denoted as CAS). Going from CAS(2,2) to CAS(6,6) lowers the gap by as much as $0.6$-$1.9$~eV. Including all $\pi$ orbitals as active, in the CASSCF(14,14) model, reduces the gaps by another $0.3$-$0.5$~eV. The $\Delta E_{\rm ST}^{\rm CASSCF(14,14)}$ values are all negative and correspond well with the $\Delta E_{\rm ST}^{sp_{\pi}}$ results obtained using CASSCF(14,14) orbitals, cf.\ Table~\ref{TAB:HF_CASSCF}. We conclude that doubly excited determinants $\{iaHL\}$, where $i<H$ and $a>L$ are $\pi$ orbitals, are the main contributors to the wavefunction,  after the leading singly excited $\{HL\}$ determinant. In Table \ref{TAB:occupations} we report sums of squares of pertinent CI coefficients for systems 2 and 4. The contribution of aforementioned double excitations is up to 15\%. This is far less than a contribution from $\{HL\}$ determinants amounting to ca.\ 70\%, but it determines ST gap inversion.
It also stresses the importance of having a small value of the HOMO-LUMO exchange integral to achieve gap inversion.

\begin{table}
\caption{S$_1-$T$_1$ energy gaps in eV obtained from CASSCF, AC0, AC$_{\rm n}$, and NEVPT2 methods for small, intermediate, and all-$\pi$ models for active spaces, compared against CC2 and Mk-MRCCSD(T) values. }
\begin{tabular}{c c S S S S c c}
\multicolumn{1}{l}{System} & \multicolumn{1}{l}{Active space} & \multicolumn{1}{l}{CAS}  & \multicolumn{1}{l}{AC0} & \multicolumn{1}{l}{ACn} & \multicolumn{1}{l}{NEVPT2} & CC2\cite{sundholm}  
 & \multicolumn{1}{l}{Mk-MRCCSD(T)} \\ \hline
\multirow{3}{*}{1} & (2,2) &  1.76 &  0.70 &  0.00 & -0.37 & \multirow{3}{*}{-0.13} & \multirow{3}{*}{-0.18} \\
                   & (6,6) & -0.18 &  0.16 &  0.01 & -0.08 &                        &                        \\
                 & (14,14) & -0.47 & -0.10 & -0.21 &  0.07 &                        &                        \\ \hline
\multirow{3}{*}{2} & (2,2) &  0.31 & -0.78 & -0.28 & -1.04 & \multirow{3}{*}{-0.24} & \multirow{3}{*}{-0.28} \\
                   & (6,6) & -0.31 & -0.02 & -0.14 &  1.01 &                        &                         \\
                 & (14,14) & -0.62 & -0.21 & -0.34 & -0.06 &                        &                         \\ \hline
\multirow{3}{*}{3} & (2,2) &  0.44 & -0.28 &  0.02 & -0.59 & \multirow{3}{*}{-0.11} & \multirow{3}{*}{-0.15}  \\
                   & (6,6) & -0.17 &  0.08 & -0.01 & -0.04 &                        &                         \\
                 & (14,14) & -0.45 & -0.05 & -0.19 &  0.09 &                        &                         \\ \hline
\multirow{3}{*}{4} & (2,2) &  0.79 & -0.45 & -0.22 & -0.92 & \multirow{3}{*}{-0.08} & \multirow{3}{*}{-0.04}  \\
                   & (6,6) &  0.18 & -0.32 & -0.01 & -0.59 &                        &                         \\
                 & (14,14) & -0.31 & -0.04 & -0.12 & -0.07$^{a}$ &                  &                         \\ \hline
\multirow{3}{*}{5}         & (2,2)                         & 0.70  & -0.64 & -0.06 & -0.94        & \multirow{3}{*}{-0.14} & \multirow{3}{*}{-0.15}        \\
                   & (6,6) & -0.12 &  0.04 & -0.33 & -0.25 &                        &                         \\
                 & (14,14) & -0.50 & -0.06 & -0.20 &  0.09 &                       &                         \\ \hline
\multirow{3}{*}{6} & (2,2) &  0.72 & -0.67 & -0.06 & -0.98 & \multirow{3}{*}{-0.12} & \multirow{3}{*}{-0.14} \\
                   & (6,6) & -0.22 &  0.08 & -0.08 & 0.05  &                        &                        \\
                 & (14,14) & -0.50 & -0.01 & -0.15 &  0.11 &                        &                        \\ \hline\hline \\
\end{tabular} 
\\
\footnotesize{$^{a}$ Due to problems with convergence of NEVPT2, the active space has been reduced to (14,13).}
\label{TAB:ST}
\end{table}

\begin{figure}[ht]
\hspace*{-1cm}

\includegraphics[width=1.0\textwidth]{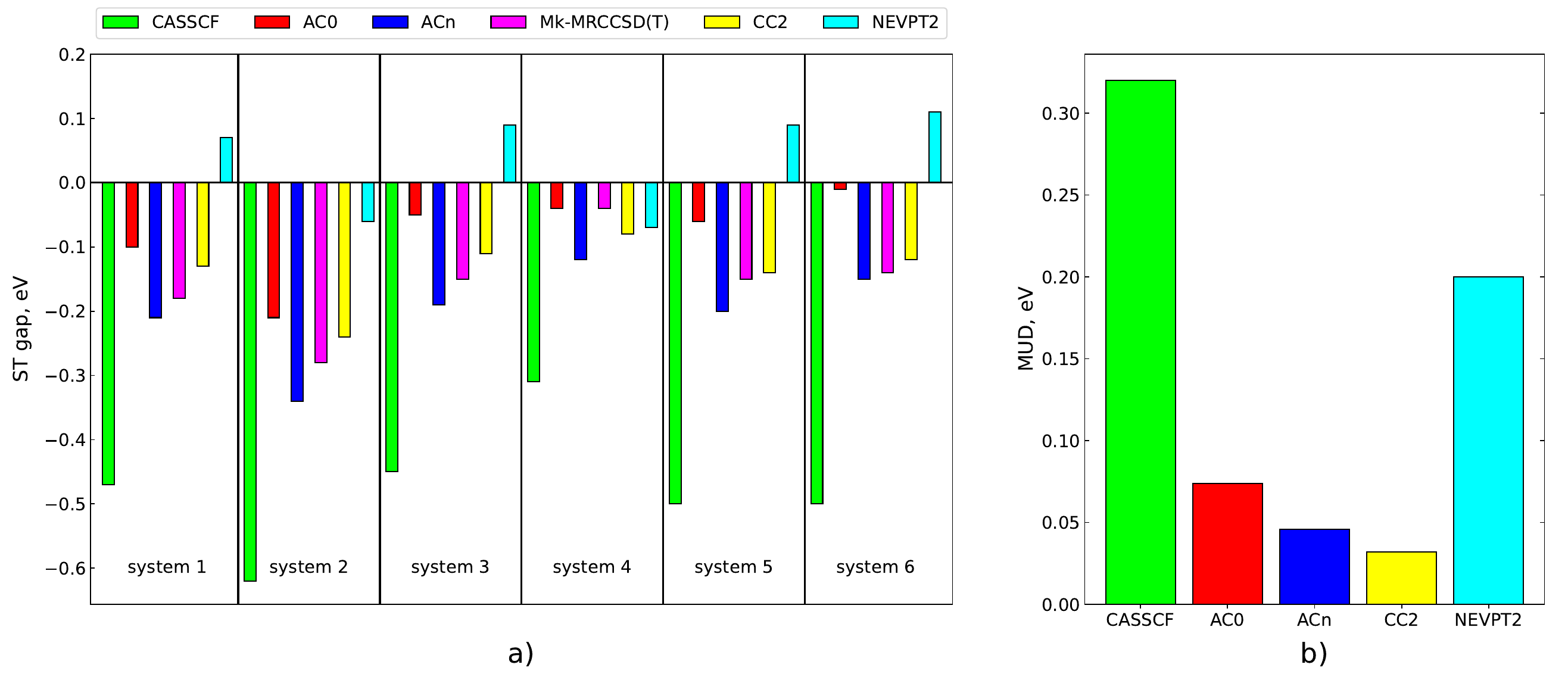}
\centering

 \caption{ S$_1-$T$_1$ energy gaps for systems 1-6, panel (a). Mean unsigned deviations of the gaps with respect to Mk-MRCCSD(T) values, panel (b). CASSCF, AC0, ACn, and NEVPT2 results correspond to (14,14) active space.} 
\label{PLOT:STgaps1to6}
\end{figure}

A comparison between CASSCF(14,14) and Mk-MRCCSD(T) reference values, cf.\ Table~\ref{TAB:ST}, reveals that CASSCF gaps are too negative with the mean unsigned deviation exceeding 0.3~eV, see also Figure \ref{PLOT:STgaps1to6}. To correct CASSCF for the missing dynamic correlation energy, we use the multireference adiabatic connection (AC) approach\cite{ac_prl,Pernal:18b,Pastorczak:18a,Pastorczak:18b,pastorczak2019capturing,Beran:21,matouvsek2023toward}, which recently has been successfully applied to predicting singlet-triplet gaps of biradicals.~\cite{Drwal:22} Two AC variants are employed in this work: AC0 and ACn. The first one is based on linearizing the adiabatic connection integrand.\cite{ac_prl,matouvsek2023toward} ACn is free of such an approximation and is expected to yield more accurate predictions.\cite{Drwal:22} It is worth mentioning that the computational cost of both AC0 and ACn scales with the $5$th power of the system size, AC0 being more efficient than ACn due to a smaller prefactor. AC methods rely only on 1- and 2-electron reduced density matrices (1- and 2-RDMs, respectively), which makes them computationally more efficient in treating large active spaces compared to canonical multireference perturbation theory methods.\cite{Beran:21}

AC0 (ACn) ST gaps are computed by adding dynamic correlation correction to the CASSCF gap according to
\begin{equation}
\Delta E_{\rm ST}^{\rm{AC}} = \Delta E_{\rm ST}^{\rm CASSCF} + E^{\rm AC}({\rm S}_1) - E^{\rm AC}({\rm T}_1) \ \ \ ,
\label{STAC}
\end{equation}
where $E^{\rm AC}({\rm S}_1)$ and $ E^{\rm AC}({\rm T}_1) $ are AC0 (ACn) adiabatic connection correlation energies computed from CASSCF  1,2-RDM's for the S$_1$ and T$_1$ states, respectively.  
Inspection of the AC0 and ACn energy gaps in Table \ref{TAB:ST} shows that accounting for dynamic correlation energy counteracts the spin polarization effect. While  spin polarization lowers the singlet state energy more than the triplet, leading to a negatively-valued ST gap, dynamic correlation shifts the S$_1$ energy upward relative to T$_1$. In the case of AC0 this effect is overestimated leading to too small gaps. AC0 gaps computed for CASSCF(6,6) change the sign back to positive for systems 1, 3, 5, and 6. Combining AC0 with CASSCF(14,14) retains the negative sign of the gaps, but their magnitudes are underestimated compared with Mk-MRCCSD(T) values, see also Figure \ref{PLOT:STgaps1to6}. 
The best accuracy is obtained if ACn is applied together with a CASSCF(14,14) model---the resulting gaps values agree up to 0.04~eV with the Mk-MRCCSD(T) reference. AC correlation energy computed for CASSCF(2,2) models yields poor results. 
This is expected, since the CASSCF(2,2) does not contain doubly excited determinants,  which AC approximations cannot amend. 
Ideally, the CASSCF model should include all of the spin polarisation, leaving only dynamic correlation energy for adiabatic connection. ACn combined with CASSCF(14,14) proves that this strategy leads to accurate predictions.

In addition to AC, in Table~\ref{TAB:ST} we present results of NEVPT2 calculations.~\cite{nevpt2} Although our previous studies show that NEVPT2 is typically as accurate as AC methods, in particular AC0\cite{ac_prl,Pastorczak:18a}, for heptazine-based systems the performance of NEVPT2 is inferior. The NEVPT2 energy gap values do not show a systematic improvement when enlarging the CAS(n,n) model. Quite contrary, the behavior is erratic and the gaps are of wrong sign, except for systems 2 and 4, even if CAS(14,14) model is employed, cf.\ Table~\ref{TAB:ST}. NEVPT2 gaps deviate on average by 0.2~eV from the reference values, see Figure~\ref{PLOT:STgaps1to6}.

The presented results show that two factors are equally important for accurate prediction of ST gap inversion: accounting for a modest contribution of double excitations and proper treatment of dynamic correlation. It is therefore not surprising that single reference methods like CC2 or ADC(2)\cite{desilva2019inverted,sundholm} are capable of predicting ST gaps as accurately as multireference approaches. The CC2 energy gaps taken from Ref.~\citenum{sundholm} (see Table~\ref{TAB:ST} and Figure \ref{PLOT:STgaps1to6}) stay in a good agreement with the multireference CCSD(T) results deviating from the latter by only $0.03$~eV. Finally, we notice that the energy gaps predicted by both single and multireference methods are quite insensitive to the basis set used, compare the results from Table \ref{TAB:ST}  and Table 2 in Supporting Information.

\section{Summary and Conclusions}
\begin{figure}[ht]
\centering
\includegraphics[width=\textwidth]{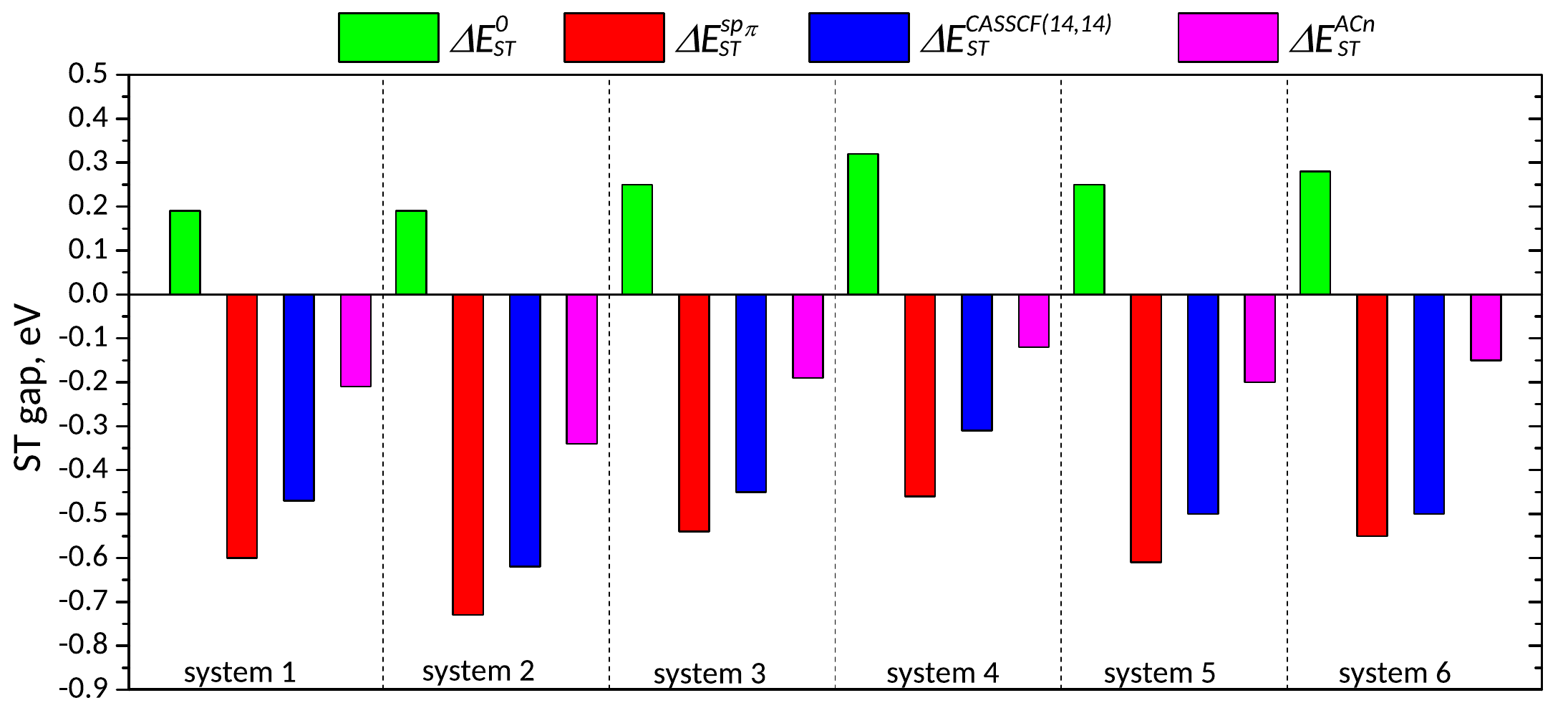}
\caption{ S$_1-$T$_1$ energy gaps for systems 1-6. $0$: no spin polarization, Eq.(\ref{E0ST}); sp${_\pi}$: PT-based model accounting for spin polarization, Eq.(\ref{STall}); ${\rm CASSCF}$: spin polarization from CASSCF(14,14); ACn: spin polarization at CAS(14,14) level and  dynamic correlation via adiabatic connection, Eq.(\ref{STAC}).}
\label{PLOT:closing_gap}
\end{figure}

We have investigated sources of S$_1-$T$_1$ energy gap inversion in heptazine-based molecules, focusing on the effect of dynamic spin polarization.~\cite{staemmler1978violation} 
We have found that spin polarization, which is equivalent to including doubly excited determinants involving HOMO, LUMO and any two of the other $\pi$ orbitals in the wavefunction expansion, drives the gap inversion.
Our findings are summarized in Figure \ref{PLOT:closing_gap}. While ignoring spin polarization leads to positive ST gaps (green bars in Figure~ \ref{PLOT:closing_gap}), accounting for this effect alone results in gaps which are too negative (red and blue bars in Figure~ \ref{PLOT:closing_gap}). For a quantitative ST gap prediction, it is essential to consider both spin polarization and accurately treat dynamic correlation energy. Dynamic correlation closes the gap, exerting an opposite effect to sp (see pink bars in Figure \ref{PLOT:closing_gap}). Comparing several multiconfigurational wavefunction approaches that account for dynamic correlation, we recommend AC-based methods as having the best accuracy/cost ratio. Among them, ACn method combined with CASSCF(14,14)  was able to reproduce accurate ST gaps for all studied heptazine derivatives.

In the investigated systems, the contribution of double excitations in wavefunctions varies from 6\% to 15\%. This relatively weak doubly-excited character of S$_1$ and T$_1$ states can be captured with single reference coupled cluster methods. Indeed, we reported a good agreement between CC2 and multireference approaches [ACn and Mk-MRCCSD(T)].

We have proposed a simple model for selecting INVEST systems which accounts for two critical factors that favor negative gaps: vanishing HL exchange integral and spin polarization. In this approach, ST gap is approximated via a PT-based expression involving orbital energies and exchange integrals. When KS-DFT orbitals are employed, the model is capable of efficient and accurate prescreening for INVEST candidates, as we have demonstrated on a test set of $\sim 100$ heptazine derivatives.  We have found a previously unknown relation between the sp contribution to the ST gap and the magnitude of the HL exchange integral, which are inversely-proportional to each other. It implies that the quantitative contribution to the gap from the effect of spin polarization is negligible if the H-L exchange interaction is relatively large.
Having the advantage of being low-cost to compute, the proposed model for the S$_1-$T$_1$ energy gap can be used in high-throughput calculations aimed at searching for INVEST systems. It can be also used to build models for the prediction of ST gap inversion by machine learning algorithms.

\section{Supporting Information}
Expressions: Hamiltonian matrix elements of doubly excited states, energy differences of doubly excited states. Tables: ST energy gaps from sp models, ST energy gaps in cc-pVDZ and cc-pVQZ basis sets, singlet- and triplet-state energies. Figures: CASSCF(14,14) orbitals for S$_1$ and T$_1$ states, ST energy gaps from sp models with HF orbitals vs. EOM-CCSD for extended set of molecules.
\section{Acknowledgement}
This work was supported by the Czech Science Foundation (Grant No. 23-04302L); the National Science Center of Poland (grant no. 2019/35/B/ST4/01310); Lodz University of Technology (Internal Grant FU2N - Fundusz Udoskonalania Umiejętności Młodych Naukowców from Excellence Initiative -- Research University);
the Charles University Grant Agency (Grant No. 218222);
the Center for Scalable and Predictive methods for Excitation and Correlated phenomena (SPEC), which is funded by the U.S. Department of Energy (DOE), Office of Science, Office of Basic Energy Sciences, the Division of Chemical Sciences, Geosciences, and Biosciences.
This work was also supported by the Ministry of Education, Youth and Sports of the Czech Republic through the e-INFRA CZ (ID:90254).

\newpage
\clearpage
\bibliography{biblio_nonhund}

\end{document}